\documentclass[prb, twocolumn, superscriptaddress]{revtex4-2}
\usepackage{amsmath,amssymb,bm}
\usepackage{hyperref}
\hypersetup{
    colorlinks=true,
    citecolor=blue,
    linkcolor=blue,
    filecolor=blue,      
    urlcolor=blue,
    pdftitle={Overleaf Example},
    pdfpagemode=FullScreen,
    }
\usepackage{graphicx}
\usepackage{epstopdf}
\usepackage{latexsym}
\usepackage[usenames, dvipsnames]{color}
\usepackage[usenames, dvipsnames]{xcolor}
\usepackage{braket}
\usepackage{float}
\usepackage[normalem]{ulem}
\usepackage{comment}
\usepackage{mathtools}
\usepackage{array}
\usepackage{tabu}
\usepackage{multirow}
\usepackage[inline]{enumitem}
\usepackage{cleveref}
\usepackage{xcolor}
\usepackage[normalem]{ulem}
\usepackage{natbib}
\usepackage{amsmath}
\usepackage{dsfont}

\begin{document}

\title{Charge modulation in the background of depleted superconductivity inside vortices}
\author{Chiranjit Mahato}
\affiliation{Indian Institute of Science Education and Research (IISER) Kolkata, Mohanpur - 741246, West Bengal, India}
\author{Anurag Banerjee}
\affiliation{Institut de Physique Théorique, Université Paris-Saclay, CEA, CNRS, F-91191 Gif-sur-Yvette, France}
\author{Catherine Pépin}
\affiliation{Institut de Physique Théorique, Université Paris-Saclay, CEA, CNRS, F-91191 Gif-sur-Yvette, France}
\author {Amit Ghosal}
\affiliation{Indian Institute of Science Education and Research (IISER) Kolkata, Mohanpur - 741246, West Bengal, India}

\begin{abstract}
We use the Bogoliubov-de Gennes (BDG) formalism to undertake a microscopic investigation of a vortex lattice in a strongly correlated, type-II, d-wave superconductor (SC) treating strong correlation within Gutzwiller formalism. We demonstrate that in the underdoped region, the vortex core changes from metallic-type to insulating-type in the presence of subdominant charge and bond order, in contrast to Mott-type, when these orders are absent. We have investigated that such subdominant order changes the structure and spectrum of the d-wave vortex in the underdoped region. We have demonstrated the formation of charge and bond modulation at the vortex center by decreasing the doping and reaching an underdoped zone.

\end{abstract}
\maketitle

\begin{figure*}[t]
    \includegraphics[width=1\textwidth]{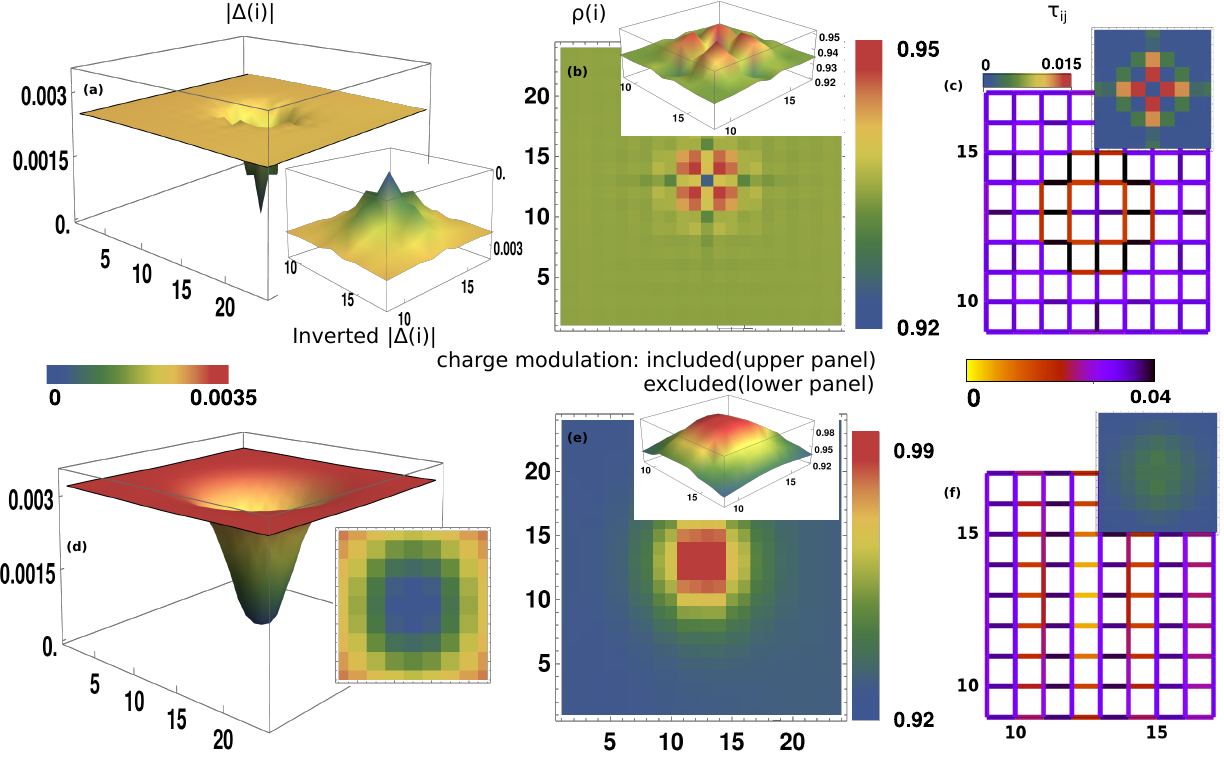}
\caption{Effect of self-consistent charge modulation at strong underdoping ($\delta=0.065$). 
(i) (a) and (d) show the spatial profile of the d-wave SC OP on a $24\times24$ magnetic unit cell. In the presence of CDW and BDW, the shape of the d-wave SC OP (a) becomes nearly conical in shape with heavily anisotropic in contrast to the flat bottom bowl shape in the absence of CDW and BDW (d)(inset in (a) and (b) shows the 3D profile of inverted d-wave SC OP and 2D profile of d-wave SC OP respectively). (ii)2D spatial profile of charge density is presented in (b) and (e). (b) shows charge density modulation at the vortex core, and (e) depicts the accumulation of charge without any modulation (in the absence of CDW and BDW order). For visual clarity, we have used two different colorbars for density. (iii)(c) and (f) show bond density $\tau_{ij}$ (where i and j are nearest neighbor lattice sites) profile. Both show a modulation in $\tau_{ij}$ at the vortex core. d-wave component of  bond density $\tau_{ij}$ is shown in the inset, shows a modulation at the vortex core in the upper panel and is absent otherwise. For visual clarity, all the figures in the inset is presented around the vortex core only (from nine to seventeen lattice spacing).  
}
    \label{f1}
\end{figure*}
\begin{figure}[t]
  \includegraphics[width=0.5\textwidth]{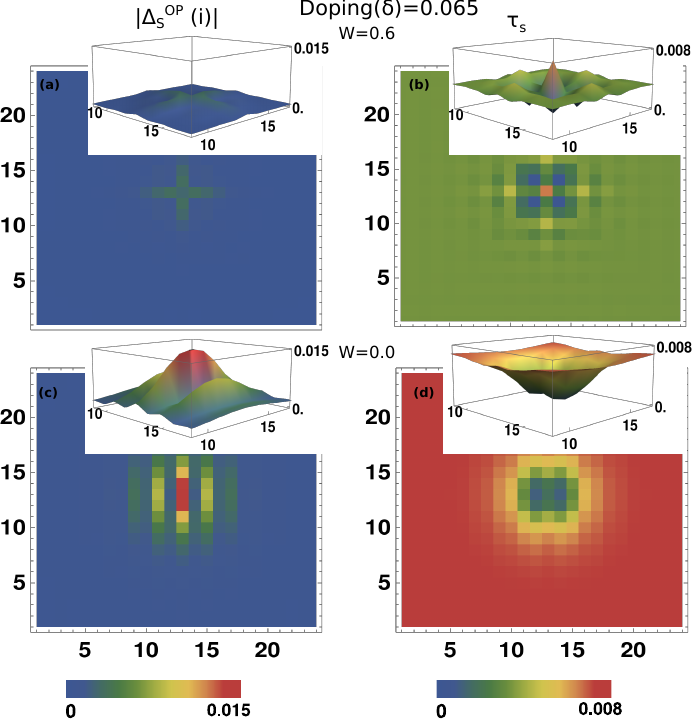}
\caption{Self-consistent solutions at $\delta=0.065$. (i) Spatial profile of the extended s-wave SC OP, $|\Delta^{OP}_s(i)|$ (a,c), on a $24\times24$ magnetic unit cell in the presence and absence of CDW and BDW, respectively. In the presence of CDW and BDW, the magnitude of the extended s-wave SC OP reduces in the strongly underdoped regions. Without CDW and BDW, $|\Delta^{OP}_s(i)|$ shows a stripe pattern near the vortex center.  (ii) 2D spatial profile of extended s-wave component of bond density $\tau_{s}$ (b,d) shows a modulation in the presence of CDW and BDW. The insets show the 3D profile of the above OPs within the vortex core (from nine to seventeen lattice spacing). 
}
    \label{f2}
\end{figure}

\begin{figure}
    \includegraphics[width=0.5\textwidth]{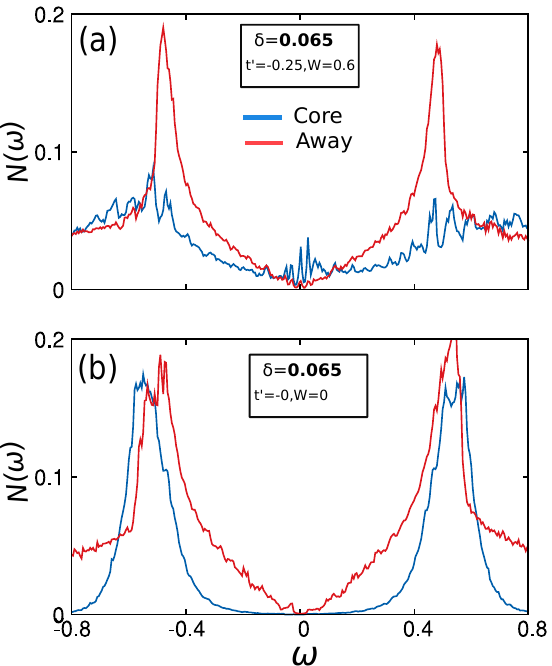}
\caption{Local density of states: LDOS near the vortex core (blue) and far away from the vortex core (red) for $\delta$=0.065 in the presence (a) and absence (b) of CDW and BDW. LDOS in the presence of CDW and BDW shows small LECS in contrast to a hard U-shaped gap in the absence of any CDW and BDW.
}   
 \label{f3}
\end{figure}

\begin{figure*}[t]
  \includegraphics[width=1\textwidth]{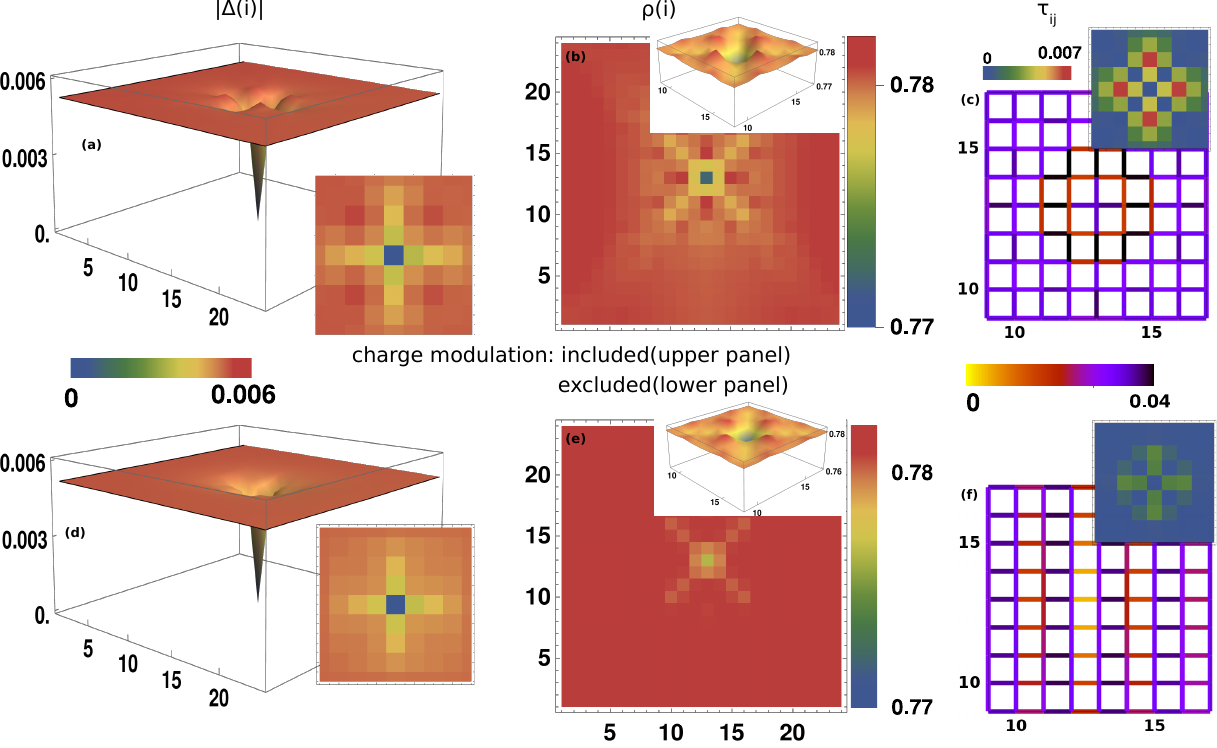}
\caption{Spatial profile for the self-consistent solutions at $\delta=0.22$. (i) Spatial profile of the d-wave SC OP, $|\Delta^{OP}_d(i)|$, on a $24\times24$ magnetic unit cell (a,d). 
ii) (b) and (e) display the 2D spatial profile of charge density. Charge density shows very weak modulation at the vortex core. (iii) Panels (c) and (f) illustrate the bond density $\tau_{ij}$ (where i and j represent nearest neighbor lattice sites) profile. Both exhibit a modulation in $\tau_{ij}$ at the vortex core. In the presence of CDW and BDW, the d-wave component of bond-density $\tau_{d}$ (insets of (c) and (f)) shows modulation at the vortex core and is absent otherwise. However, the magnitude is smaller compared to the strong underdoped case. For better visual clarity, all the inset figures are focused solely on the vortex core, covering a range from nine to seventeen lattice spacings.    
}
    \label{f4}
   \end{figure*} 
\begin{figure}[!htb]
    \includegraphics[width=0.5\textwidth]{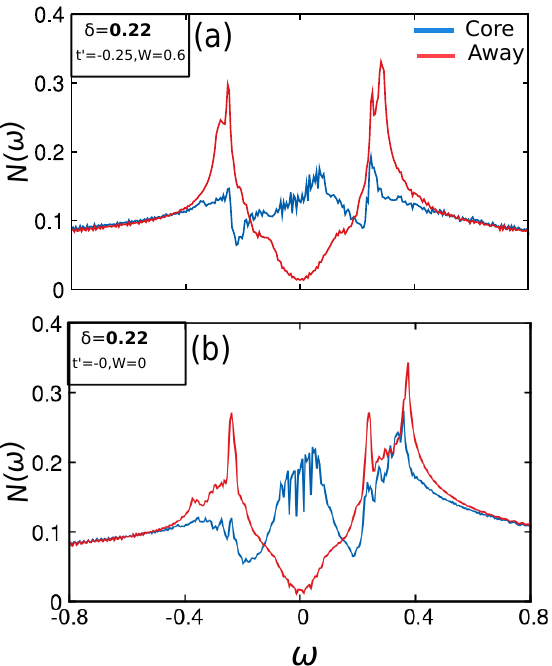}
\caption{ Local density of states: LDOS near the vortex core (blue) and far away from the vortex core (red) for $\delta$=0.22 (a,b). Both in the presence and absence of CDW and BDW, LDOS at the vortex core shows LECS. LDOS away from the vortex core shows d-wave SC gap.
}   
 \label{f5}
\end{figure}

\textit{Introduction: \textemdash } Vortices are topological defects, and they arise as low-lying excitation of superconductors, particularly when exposed to external orbital magnetic fields~\cite{kopnin2001theory}.
Unbinding of vortex-antivortex pairs was put forward as the mechanism of a thermal melting in 2D by Kosterlitz and Thouless~\cite{JMKosterlitz_1973,RevModPhys.89.040501} -- a case for which the usual order-disorder transition does not occur among condensed phases. Thus, vortices attracted research attention from early on, and their understanding remains crucial in the wake of comprehension of topology in condensed phases of matter. In conventional superconductors, the magnetic field creates a regular pattern of Abrikosov vortices~\cite{ABRIKOSOV1957199,kopnin2001theory}. Each vortex has a normal metallic core with a diameter of $\xi$ with circulating currents around the vortex on the scale of the penetration depth $\lambda$~\cite{Tinkham}. The number of vortices goes up as field H goes up. When the critical field $H_{c2}$ is reached, pairing amplitude is suppressed everywhere by overlapping cores, and the superconductor transitions into a metal~\cite{ABRIKOSOV1957199}.  

Unconventional superconductors, particularly cuprate superconductors where strong electronic repulsion germinates superconductivity with a d-wave pairing symmetry~\cite{PhysRevB.107.L140505,PhysRevLett.87.217002,PhysRevB.66.094513, RevModPhys.78.17,RevModPhys.79.353}, promotes many surprises when contrasted with weak-coupling d-wave superconductivity (dSC)~\cite{PhysRevB.107.L140505,PhysRevB.52.R3876,PhysRevB.68.012509}. On the other hand, the role of strong correlations is not limited to the weak-coupling picture of dSC alone; it is the same strong electronic correlations that give rise to subdominant yet competing broken-symmetry orders, such as charge modulation  orders~\cite{PhysRevB.105.134505,PhysRevB.97.174511,PhysRevResearch.5.033028}. Such additional orders compete with dSC, particularly at underdoping, where the effects of strong correlations are paramount due to the proximity of the Mott-insulator. The subtle interplay among the competing ordered ground states is believed to induce the amazements of unconventional superconductivity~\cite{RevModPhys.87.457,doi:10.1126/science.1107559}.

What role do such competing orders play in shaping the physics of vortices in a strongly correlated d-wave superconductor? Addressing this precise question, we develop a comprehensive picture in this manuscript.

 Charge-density waves (CDWs), which can exhibit either unidirectional or bidirectional patterns in two-dimensions (2D), have been observed in many hole-doped high-temperature superconductors (HTSCs)~\cite{RevModPhys.75.1201,doi:10.1146/annurev-conmatphys-031115-011401}. At low temperatures ($T$), these CDWs remain short-ranged and are suppressed by the onset of superconductivity~\cite{doi:10.1126/science.1223532,PhysRevLett.110.137004,PhysRevB.90.054513}. Magnetic field amplifies both the intensity and spatial extent of the CDW ordering~\cite{wu2011magnetic,hamidian2016atomic,zhou2017spin,wen2023enhanced}. The study of modulated charge order on the vortex state of a d-wave superconductor gathered momentum~\cite{hoffman2002,PhysRevResearch.5.033028,wu2011magnetic,Wu2013,Tu2016} since Hamidian and collaborators~\cite{doi:10.1126/science.aat1773} found the emergence of compelling signatures of field-induced pair density wave (PDW) with a periodicity of eight lattice spacing (8$a_{0}$), within the `halo' region surrounding a vortex in underdoped cuprate, coexisting with CDW order. The emergence of charge ordering is not just contingent on the depletion of SC order; there are recent experiments proposing charge order to persist for extreme underdoping coexisting with antiferromagnetism, and moreover, they are thought to germinate superconductivity~\cite{Zou2024}. 

   Several theoretical~\cite{PhysRevResearch.5.033028,PhysRevB.97.174510,PhysRevB.97.174511} attempts have been made to unfurl the role of such modulated charge orders in the mixed state of an unconventional superconductor. It is only prudent to build in the possibility of the emergence of subdominant charge orders in the vortex core where even the unconventional superconductivity must be depleted in the mixed state. This opens up the stage for studying the abstruse interplay of Mottness-induced dSC and the competing orders -- both originating from strong electronic correlations! Unfortunately, theoretical techniques are limited to addressing such an outstanding problem, making it difficult for convergence to materialize. We treat this problem using a simple yet intuitive methodology by numerically implementing it for systems with broken translation symmetry, as we outline below.

  Describing the underdoped cuprates by the $t$-$J$ Hamiltonian and having analyzed it within a fully self-consistent scheme of Gutzwiller's renormalized mean-field theory, referred to as GIMT, our key results are the following: (a) Emergence of charge and bond order modulation is found at the vortex core, specifically at strong underdoping. The intensity of these modulations decreases as we increase the doping towards optimal to overdoped region. (b) In the underdoped region, the presence of subdominant charge and bond order alters the nature of the vortex core from a Mott-type insulator to a charge-ordered insulator. (c) The structure and spectrum of the vortex modify, reflecting the presence of local subdominant orders. (d) The profile of the local density of the state (LDOS) shows traces of low-energy core states (LECS) -- the extent of low-lying LDOS is distinct from both a Mott-core featuring a hard U-shaped gap~\cite{PhysRevB.107.L140505} and also from a weak-coupling signature of a broad hump~\cite{PhysRevB.52.R3876} resulting from quasi-bound core states~\cite{PhysRevLett.80.4763}.
  (e) In the underdoped region, the presence of charge and bond order suppresses other pairing symmetries, such as extended s-wave superconducting order and bond-density wave order.

\textit{Model \& method: \textemdash } We consider the t-J model~\cite{K_A_Chao_1977, PWAnderson_2004}, which is derived from a Hubbard model with strong onsite repulsion ($U\gg t$) as an effective low energy Hamiltonian. In addition, we include an additional repulsion $W$ on next nearest neighboring sites of the lattice. Thus, the Hamiltonian is given by,
\begin{equation}
  \begin{split}
  {\cal H}_{t-J-W}=\sum_{{<ij>}{\sigma}}\hat{P}\left(t_{ij}e^{i \phi_{ij}} \hat c{^\dag_{i \sigma }} \hat c{_{j \sigma }} + h.c. \right)\hat{P}  
  +\sum_{i}(-\mu ){\hat n_{i}} \\
  + J \sum_{<ij>} \hat{P}\left ( {\bf S_{i}. S_{j}} -\frac{\hat{n}_{i}.\hat{n}_{i}}{4}                 \right )\hat{P} 
  +W\sum_{<ij>,\sigma}\hat{P}(\hat{n}_{i\sigma}\hat{n}_{j\sigma} )\hat{P}
  \end{split}
\end{equation}
Here, $W$ can be thought of as the remnants of a screened Coulomb repulsion, but for us, it acts as a knob to tune a charge order independent of superconductivity.
Also, $c{^\dag_{i \sigma }}$ (${c_{i \sigma }}$) creates (annihilates) an electron on site $i$ with spin ${\sigma}$ on a two-dimensional (2D) square lattice, and $n_{i}$ denotes electron-density operators.
 $t_{ij}=-t$ if $i,j$ are nearest neighbours (denoted as $\langle ij\rangle$) and  $t_{ij}$=$t'$ if $i,j$ are the next nearest neighbour sites. We fix $t=1$,$t'$=0.25t, and set all energies in units of $t$. 
${\bf S_{i}}$ is the spin operator and the exchange interaction $J=4t^2/U$~\cite{PhysRevB.52.615}. We choose $U=12t$ and $W=0.6t$ for all our results reported here.
 $\mu $ is the chemical potential that fixes the average density, $\rho$, of the system. The orbital magnetic field enters into the system in the form of Peierls factor: 
  $\phi_{ij}=\frac{\pi}{\phi_0}\int^{j}_{i}A.dl$
  where $\phi_0=\frac{hc}{2e}$  is the superconducting flux quantum, and A is the vector potential in Landau gauge  $A = Hx\hat {y}$ which ensures a uniform orbital field aligned along the z-axis, i.e.,$ H = H\hat{z}$. \\
Strong onsite repulsion prohibits double occupancy in the Hilbert space, and $\hat{P}=\Pi_i \left(\mathds{1}-\hat{n}_{i\uparrow}\hat{n}_{i\downarrow}\right)$ projects out all double occupancy. These restrictions are implemented via Gutzwiller approximation (GA)~\cite{PhysRevB.76.245113}, which amounts to renormalized the parameters $t$ and $J$ of ${\cal H}_{t-J-W}$ by local density-dependent factors, known as  Gutzwiller renormalization factors (GRFs)~\cite{PhysRevB.76.245113,FCZhang_1988}, such that they mimic the projection due to strong repulsion. Restricted hoping, on average, reduces $t_{ij}$ due to the double occupancy prohibition, whereas the effective exchange (and hence $J_{ij}$) increases because of enhanced overall single occupancy~\cite{PhysRevB.76.245113}.

Here, we work with a magnetic unit cell containing two superconducting (SC) flux quanta -- an even number of SC flux quanta is necessary in our system for implementing periodic boundary conditions. The above ${\cal H}_{t-J-W}$ is solved self consistently~\cite{PhysRevB.63.020505,Chakraborty_2014} on a magnetic unit cell and then extending the wave function on a system typically containing $16\times8$ unit cells using repeated zone scheme (RZS). We have compared the Gutzwiller inhomogeneous mean-field theory (GIMT) results for two values of $W$. When $W=0$, the situation is analogous to ~\cite{PhysRevB.107.L140505}, and the self-consistent solution allows vortices in a d-wave superconducting (dSC) ground state. We consider $W=0.6$ that permits self-consistent modulation of charge order (both on sites as well as on bonds), particularly at the vortex core where dSC is depleted.
\textit{Results: \textemdash} Inclusion of nearest neighbor repulsion, as well as next-nearest neighbor hoping in a $t-J$ Hamiltonian help generates CDW and BDW order in the absence of vortices~\cite{PhysRevB.105.134505} (arising from the orbital magnetic field). The vortex lattice of a strongly correlated superconductor that excludes charge modulation order is shown to generate Mott insulating vortex cores at strong underdoping while featuring usual metallic vortex core at optimal to overdoping~~\cite{PhysRevB.107.L140505}. However, in such calculations, the effects of strong electronic repulsion were limited to how the exclusion of double occupancy (arising from strong repulsion) modifies dSC pairing. On the other hand, there are mounting pieces of evidence that the proximity of Mottness in underdoped cuprates generates charge modulation orders~\cite{doi:10.1073/pnas.2302099120,doi:10.1126/science.1243479,Cai2016} -- the modulation wave-vector being sensitive to various tuning parameters~\cite{PhysRevB.105.134505,PhysRevResearch.5.033028,choubey2017incommensurate}. Thus, the study of a vortex lattice of a strongly correlated underdoped dSC must accommodate for charge modulation order self-consistently -- in particular, such ordering might get stabilized at the vortex cores where dSC is depleted. Thus, it necessitates the inclusion of these additional low-energy broken symmetry orders, in addition to suppressing double occupancy in a Gutzwiller projected inhomogeneous mean-field theory (GIMT). 
In the following, we compare how dSC, charge modulation, and bond density modulation in and around the vortex core of a strongly correlated dSC alter in the presence of $W$ and $t^{\prime}$, which seed charge modulation. We start discussing our results by presenting dSC pairing amplitude:
$\langle\tilde{c}_{i\sigma}\tilde{c}_{j\bar{\sigma}}\rangle_{\psi} \approx g^{t}\Delta_{ij}$~\cite{PhysRevB.78.115105,PhysRevB.96.134518}. Here $\langle \cdots \rangle_{\psi}$
denotes the expectation value, evaluated using GIMT formalism. From now on, we refer to the aforementioned situations as Case (1): Calculations that allow self-consistent CDW and BDW orders, in addition to dSC pairing (W=0.6, $t^{\prime}=-0.25$).
Case (2): Calculations that do not allow self-consistent generation of CDW and BDW on top of dSC order (W=0, $t^{\prime}= 0$).
\begin{center}
 \textit{\bf{Strongly underdoped region}}
 \end{center}

\textit{d-wave SC order: \textemdash}

The local dSC order parameter, in terms of corresponding bond variables is defined as:
\begin{eqnarray}
\Delta^{OP}_d(i) &=& \frac{J}{4} \left\vert [g^{t}_{i,i+\hat{x}}\Delta^{\hat{x}}(i)+g^{t}_{i,i-\hat{x}}\Delta^{-\hat{x}}(i)\right. \nonumber \\ 
&-& \left. e^{ibx}g^{t}_{i,i+\hat{y}}\Delta^{\hat{y}}(i)-e^{-ibx}g^{t}_{i,i-\hat{y}}\Delta^{-\hat{y}}(i)]\right\vert
\label{Eq:DelOP}
\end{eqnarray}
(here $b=H/\phi_0$) and is shown in Fig.~\ref{f1} (a,d) for $\delta=0.065$, which corresponds to strong underdoping. Only half of the magnetic unit cell containing a single dSC vortex is shown. We present results from cases (1) and (2) (defined already) in upper and lower panels to highlight their contrast.
Calculations using the protocol of case (2) found that self-consistent vortices assume the shape of a flat-bottom bowl as the system turned strongly underdoped instead of the nearly doping-independent conical one within the weak-coupling theory (See Appendix). Such a change was attributed to emerging Mottness due to its proximity to a Mott insulator. In contrast, our results in Fig.~\ref{f1}(a) show that the competing charge modulation orders in the ground state deform the shape in a non-trivial way. While the flat-bottom bowl goes back partially towards a conical-like shape, but more importantly, the shape of the vortex core becomes highly anisotropic (shown in the {\it inverted}  form as an inset to Fig.~\ref{f1} (a) for clarity) in response to the spatial variation of charge density (discussed below). We emphasize that the topology of the vortices upon allowing charge order is quite distinct from GIMT~\cite{PhysRevB.107.L140505} and IMT findings (shown in the Appendix). 
\textit{Local charge density : \textemdash}
To gain a deeper understanding of the nature of the spatial profile of the pairing amplitude discussed above, we proceed to examine the local charge density, which is depicted in Fig.~\ref{f1} (b,e), by including self-consistent charge modulations and excluding them, respectively. While the local charge density undergoes a minor depletion, irrespective of the average electron density in the system within a weak-coupling description (See Appendix for IMT results). In contrast, GIMT calculation at strong underdoping allowing only dSC as the sole self-consistent order yielded a large accumulation of electrons in the vortex core (Fig.~\ref{f1}(e)), reaching near half-filling -- the maximal possible value! This was attributed to the emergence of a Mott-insulating vortex core due to the enhanced role of strong correlations as $\delta \rightarrow 0$. Accommodating charge ordering, our present calculation opens up the possibility of a subtle interplay of Mottness that favors a half-filling in the core region and spatial variation of electronic density arising from local charge modulation ordering. The resulting self-consistent density reflects a rich spatial profile, as seen from  Fig.~\ref{f1}(b). Together with Fig.~\ref{f1}(a), Fig.~\ref{f1}(b) put forward our main results. The charge modulation in the vortex core region, from our calculation, shows an overall electron accumulation in the vortex core relative to the background, the very nature of the modulation, as well as the strong depression of the charge density at the vortex center, also indicates that the vortex cores are of non-Mott type, contrary to what was found earlier~\cite{PhysRevB.107.L140505}.

\textit{Bond-density : \textemdash}
Recent experiments strongly indicated that along with the spatial modulation in electronic site density, the bond density, $\tau_{j\sigma}(i) = \langle {\hat c}^{\dagger}_{i\sigma} {\hat c}_{j\sigma}\rangle$, where $i,j$ are nearest neighboring sites, also shows a non-trivial pattern in the ground state of strongly correlated cuprate superconductors. In the absence of any local or global magnetic ordering, which we consider here, $\tau_{\sigma}\equiv\tau_{-\sigma}$. Results in Fig.~\ref{f1}(c,f) demonstrate the variation of bond density ($\tau_{ij}$), where i,j are the nearest neighboring sites. The d-wave component of bond density is extracted from $\tau_{ij}$. The bond density with a d-wave form-factor, $\tau_{d}(i)$, is defined in terms of $\tau_{j(i)}$ in a manner analogous to $\Delta^{OP}_d(i)$:
\begin{gather*} 
\begin{split}
\tau_{d}(i)=\frac{1}{4}\left(\tau_{x}(i)+\tau_{-x}(i)-\tau_{y}(i)-\tau_{y}(i)\right)
\end{split}
\end{gather*}
The spatial profile of $\tau_{j(i)}$ is presented on the bonds for doping $\delta=0.065$ in Fig.~\ref{f1}(c) for cases with $W=0.6t$ and $t^{\prime}=-0.25t$, whereas, both these parameters are set to zero for results in Fig.~\ref{f1}(f), and the corresponding insets show $\tau_{d}(i)$. For visual clarity, All panels of Fig.~\ref{f1} focus only near the vortex core -- the spatial fluctuations of order parameters beyond these regions are strongly suppressed.

Our results indicate that a strong correlation also makes the bond density inhomogeneous in the vortex core region. The profile of $\tau_{ij}$ shows variation over a smaller radius in Fig.~\ref{f1}(c) when CDW and BDW are treated self-consistently than when they are neglected in Fig.~\ref{f1}(f). On the other hand, the strength of $\tau_{d}(i)$, shown in the insets, is larger in our new results in Fig.~\ref{f1}(c) than when non-selfconsistent CDW and BDW are considered in Fig.~\ref{f1}(f).

\textit{Extended s-wave ordering:\textemdash}
While our model with chosen parameters generates self-consistent pairing amplitude $\Delta_{ij}$ as well as the density on the bonds $\tau_j(i)$ such that when converted to site variables, only their corresponding d-wave components, such as $\Delta^{OP}_d(i)$ and $\tau_d(i)$, survive and the respective extended s-wave components (defined below) vanishes for the uniform superconductor (in the absence of vortices, i.e, $H=0$). These extended s-wave components are defined as:
\begin{gather*} 
\begin{split}
\Delta^{OP}_{xs}(i)=(J/4)|[g^{t}_{i,i+\hat{x}}\Delta^{\hat{x}}(i)+g^{t}_{i,i-\hat{x}}\Delta^{-\hat{x}}(i)+\\e^{ibx}g^{t}_{i,i+\hat{y}}\Delta^{\hat{y}}(i)+e^{-ibx}g^{t}_{i,i-\hat{y}}\Delta^{-\hat{y}}(i)]|
\end{split}
\end{gather*}
for the extended s-wave pairing amplitude (here $b=H/\phi_0$), and the corresponding component of bond density is:
\begin{gather*} 
\begin{split}
\tau_{xs}(i)=0.25(\tau_{x}(i)+\tau_{-x}(i)+\tau_{y}(i)+\tau_{y}(i))
\end{split}
\end{gather*}
where $\tau_{ij}=\sum_{\sigma}\langle c^{\dagger}_{i \sigma}c_{j \sigma}\rangle$ , and $i$, $j$ are nearest neighboring sites. Even though $\Delta^{OP}_{xs}(i)$ and $\tau_{xs}(i)$ are absent in a uniform system for our choice of parameters, the vortex lattice arising from the applied orbital magnetic field breaks the translation symmetry and thereby brings these orders into life -- particularly at the vortex cores where the standard dSC order depletes. Such additional pairing symmetries have also been observed~\cite{Hamidian2016,Achkar2016,McMahon,doi:10.1126/sciadv.aaz1708}. The spatial profiles of $\Delta^{OP}_{xs}$ and $\tau_{xs}$ from our calculations are presented in Fig.~\ref{f2}((a,c) and (b,d)) respectively. While these additional orders germinate at the vortex core, we find that their intensity remains far weaker than the corresponding d-wave components.  The nature and shape of the spatial profiles of $\Delta^{OP}_{xs}$ and $\tau_{xs}$ differ between the two cases: With dSC as the sole self-consistent order of our strongly correlated Hamiltonian, a stripe pattern of $\Delta^{OP}_{xs}$ is found around vortex core, whereas a flat-bottom bowl shape of $\tau_{xs}$ is realized. Self-consistent generation of charge modulations not only kills the stripe pattern of $\Delta^{OP}_{xs}$ but also suppresses its magnitude. It also affects the $\tau_{xs}$, producing a modulated pattern in the core region and with its peak at the vortex center.

\begin{figure}[h]
  \includegraphics[width=0.5\textwidth]{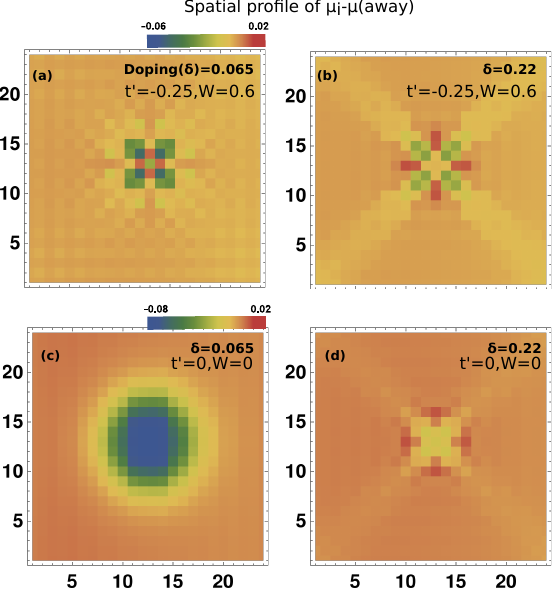}
\caption{Effective chemical potential: Spatial variation of the effective chemical potential around a vortex core, at $\delta$=0.065(a,c) and $\delta$=0.22(b,d). $\mu$ (away) is the effective chemical potential away from the vortex core. For $\delta$=0.065, in the presence (a) and absence (c) of CDW and BDW order, $\mu_{i}$ - $\mu$ shows modulation and a prominent dip in local potential, respectively. For $\delta$=0.22, $\mu_{i}$ - $\mu$ exhibits small modulation in local potential distribution around the vortex core in the presence and absence of CDW and BDW order.  } 
    \label{f6}
   \end{figure}

   \begin{figure*}[t]
 \includegraphics[width=1\textwidth]{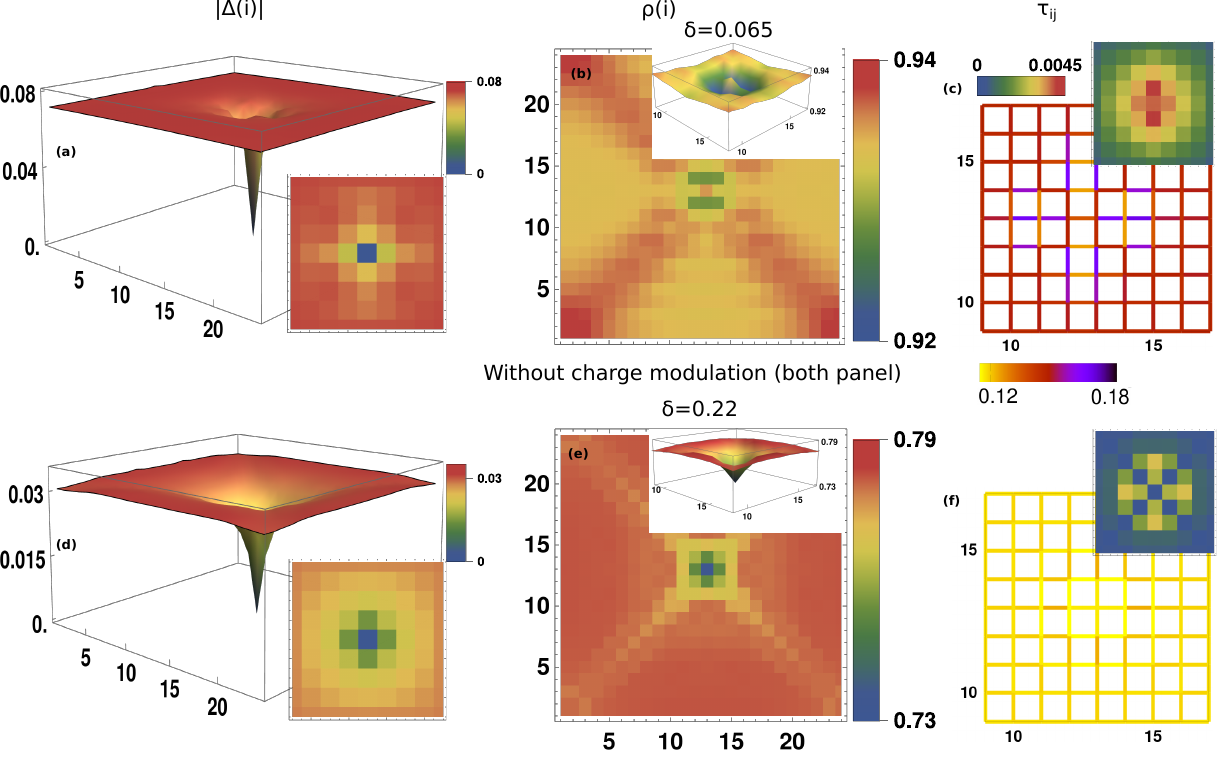}
\caption{Spatial profile for the self-consistent solutions without charge modulation at $\delta=0.22$ and  $\delta=0.065$  .(i) Spatial profile of the d-wave SC OPA, $|\Delta^{OP}_d(i)|$, on a $24\times24$ magnetic unit cell (a,d). Both are regular conical in shape. 
ii) (b) and (e) display the 2D spatial profile of charge density. Charge density shows a very weak dip around the vortex core. (iii) Panels (c) and (f) depict the profile of the bond density $\tau_{ij}$ (where i and j refer to nearest neighbor lattice sites). Both panels show no modulation in $\tau_{ij}$ at the vortex core.  d-wave component of bond-density $\tau_{d}$ (insets of (c) and (f)) shows very weak modulation at the vortex core. The inset figures are displayed around the vortex core (form nine to seven lattice spacing) to enhance visual clarity.    
}
    \label{f7}
   \end{figure*} 

  \textit{Local density of states (LDOS):\textemdash}
 The LDOS at the vortex core has been under experimental and theoretical scanners for quite some time. Early experiments provide compelling evidence for sharp zero-bias peak signaling bound state in the LDOS at the vortex core of a conventional SC~\cite{HESS1991422} as predicted by Caroli, deGenes, Matricon~\cite{CAROLI1964307}. However, for weakly coupled d-wave SC~\cite{PhysRevB.52.R3876}, the sharp peaks become broader due to the formation of quasi-bound vortex states as the d-wave gap closes in the nodal directions. In contrast, such LECS was not observed in optimal to the underdoped region of high $T_{c}$ cuprates~\cite{PhysRevLett.75.2754, HOOGENBOOM2000440, PhysRevLett.85.1536} and recently, the signature of LECS has been found for strongly overdoped region~\cite{PhysRevX.11.031040}. Various efforts have been made to understand the absence of LECS, with the lack often attributed to strong electronic correlations in these materials, which tend to generate competing orders. In the following section, we explore how local charge modulations in vortex cores in strongly correlated dSC modify the low energy spectrum of the vortex core.

Our findings are depicted in Fig.~\ref{f3}(a,b) respectively for our cases (1) and (2).

Within GIMT formalism, LDOS is expressed as:
\begin{equation}
N(i,\omega)=\frac{1}{N}\sum_{k,n}g^t_{ii}[|u^k_{n}(i)|^2\delta(\omega - E_{k,n})+|v^k_{n}(i)|^2\delta(\omega + E_{k,n})]
\end{equation}
Here $\{ u^k_{n}(i),v^k_{n}(i)\}$ are the local Bogoliubov wave functions, $E_{k,n}$ corresponding to energy eigenstates, and N is the total number of eigenstates.

With the inclusion of competing orders, the nature of the vortex core changes from a Mott-type featuring nearly a hard gap in LDOS at the vortex core to a non-Mott type for which the corresponding LDOS has a significant weight of low energy core states (LECS). This contrast is highlighted in Fig.~\ref{f4}(a,b) from our results at strong underdoping.

\begin{center}
 \textit{\bf{Results in optimal to overdoped region}}
 \end{center}
So far, we discussed our results by contrasting our findings obtained from case (1) and case (2).
However, the comparisons were kept limited only to strong underdoping. The reason is two-fold: (a) The competing charge orders are found at the underdoped regime in cuprate superconductors~\cite{Chang2012,hoffman2002,Wise2008,Frano_2020} -- a typical case of a strongly correlated dSC. (b) Our GIMT technique relies on full suppression of double occupancy, which aptly renders the system strongly correlated in the limit of strong underdoping, where such contrasts are discernible.
However, similar comparisons were also carried out for the optimal to the overdoped region. As anticipated, the degree of contrast decays as we march towards overdoping. An illustration, similar to Fig.~\ref{f1} and Fig.~\ref{f3} for a representative comparison is presented for $\delta=0.22$ in Fig.~\ref{f4} and Fig.~\ref{f5}. These establish the insignificance of charge ordering due to the subdued effects of electronic correlations at a parameter regime far from Mottness. The extended s-wave component of superconductivity, as well as the charge and bond ordering at the vortex core, is negligible in this case. Such a trend in the doping-dependence has been verified for a few other intermediate doping values.

  \textit{Effective chemical potential:\textemdash}
As emphasized above, the stark differences in different observable from the two calculations arise chiefly from the reorganization of the local density in the vortex core. This involves the accumulation of electronic density in a calculation that neglects charge orders and considers dSC alone at strong underdoping, which gets smeared out in a non-trivial way when such competing orders are included. It is natural to ask how a mean-field theory captures such effects. To address this, we produce in Fig.~\ref{f6} the profile of the effective local chemical potential, i.e. $\mu_{i}$ - $\mu({\rm away})$ on a magnetic cell containing a superconducting vortex (in a vortex lattice). Note that the effective chemical potential contains all information on Gurzwiller-renormalized Hartree shifts. The results are presented for $\delta=0.22$ and $0.065$, for the two cases with $W=0.6$ in Fig.~\ref{f6}(a,b) and for $W=0$ Fig.~\ref{f6}(c,d). In a strongly underdoped region ($\delta=0.065$), allowing charge order causes $\mu_{i}$ to self-consistently generate a weak dip at the center of the vortex core along with its modulation in the region surrounding the core. The local density responds to this effective potential and produces its spatial profile presented in Fig.~\ref{f1}(b). The overdoped region ($\delta=0.22$) also produces variations in the effective local chemical potential, albeit with weaker modulations, which was reflected in Fig.~\ref{f4}(b).
In the strongly underdoped region ($\delta=0.065$), $\mu_{i}$ self-consistently generates a prominent dip near the vortex core for $W=0$, causing electrons to accumulate locally. This, in turn, enhances the local density towards unity (See Fig.~\ref{f1}(e)). In optimal to the overdoped region ($\delta=0.22$), $\mu_{i}$ shows a very weaker dip near the center of the vortex, resulting in a local density as depicted in Fig.~\ref{f4}(e).

\textit{Conclusion:\textemdash}
  We illustrate how the structure of SC OP and the spectrum of a strongly correlated d-wave vortex get revised due to the presence of subdominant modulating charge and bond order. We have looked at the fact that the vortex core, in the presence of charge and bond order, becomes metallic-type to insulating as it approaches the underdoped region, in contrast to the Mott-insulating gap in the underdoped region, where no such dominant order exists. Findings from effective chemical potential and LDOS calculations suggest the establishment of charge and bond order towards the vortex center.

 \textit{Acknowledgement.} All authors acknowledge research funding from CEFIPRA (Grant No. 6704-3).

\appendix
\centerline{\textbf{Appendix}}
To effectively emphasize the impact of charge and bond modulation on the structure of dSC and local charge density, we include in the appendix the results from the IMT calculations that ignore the strong correlation effect.

\section{Order parameters}
\label{ORDER PARAMETERS (OP):}
Below, we discuss the spatial profile of d-wave superconducting (dSC) Order parameter amplitude (OPA), local charge density and bond density to contrast the results obtained from self-consistent charge order modulation.

\textit{dSC order parameter amplitude:\textemdash}
 Fig.~\ref{f7}(a,d) shows the spatial profile of dSC OPA for $\delta=0.22$ and  $\delta=0.065$. In contrast to the result presented in Fig.~\ref{f1}(a,d) in the main text, it shows a regular conical shape.

\textit{Local charge density:\textemdash}
 Fig.~\ref{f7}(b,e) shows the spatial profile local charge density for $\delta=0.22$ and  $\delta=0.065$. In IMT calculation, irrespective of doping, local charge density shows a weak dip around the vortex core in contrast to the result presented in Fig.~\ref{f1}(b,e) in the main text.

\textit{Bond density:\textemdash} We have presented our results on bond density, $\tau_{j\sigma}(i) = \langle {\hat c}^{\dagger}_{i\sigma} {\hat c}_{j\sigma}\rangle$, where $i,j$ are nearest neighboring sites and the bond density with a d-wave form-factor, $\tau_{d}(i)$ in the inset of Fig.~\ref{f7}(c,f) for doping $\delta=0.22$(c) and  $\delta=0.935$(f). As expected, in the absence of a strong correlation, these orders show a very weak modulation around the vortex core.

\bibliography{ref}

\begin{thebibliography}{59}%
\makeatletter
\providecommand \@ifxundefined [1]{%
 \@ifx{#1\undefined}
}%
\providecommand \@ifnum [1]{%
 \ifnum #1\expandafter \@firstoftwo
 \else \expandafter \@secondoftwo
 \fi
}%
\providecommand \@ifx [1]{%
 \ifx #1\expandafter \@firstoftwo
 \else \expandafter \@secondoftwo
 \fi
}%
\providecommand \natexlab [1]{#1}%
\providecommand \enquote  [1]{``#1''}%
\providecommand \bibnamefont  [1]{#1}%
\providecommand \bibfnamefont [1]{#1}%
\providecommand \citenamefont [1]{#1}%
\providecommand \href@noop [0]{\@secondoftwo}%
\providecommand \href [0]{\begingroup \@sanitize@url \@href}%
\providecommand \@href[1]{\@@startlink{#1}\@@href}%
\providecommand \@@href[1]{\endgroup#1\@@endlink}%
\providecommand \@sanitize@url [0]{\catcode `\\12\catcode `\$12\catcode
  `\&12\catcode `\#12\catcode `\^12\catcode `\_12\catcode `\%12\relax}%
\providecommand \@@startlink[1]{}%
\providecommand \@@endlink[0]{}%
\providecommand \url  [0]{\begingroup\@sanitize@url \@url }%
\providecommand \@url [1]{\endgroup\@href {#1}{\urlprefix }}%
\providecommand \urlprefix  [0]{URL }%
\providecommand \Eprint [0]{\href }%
\providecommand \doibase [0]{https://doi.org/}%
\providecommand \selectlanguage [0]{\@gobble}%
\providecommand \bibinfo  [0]{\@secondoftwo}%
\providecommand \bibfield  [0]{\@secondoftwo}%
\providecommand \translation [1]{[#1]}%
\providecommand \BibitemOpen [0]{}%
\providecommand \bibitemStop [0]{}%
\providecommand \bibitemNoStop [0]{.\EOS\space}%
\providecommand \EOS [0]{\spacefactor3000\relax}%
\providecommand \BibitemShut  [1]{\csname bibitem#1\endcsname}%
\let\auto@bib@innerbib\@empty
\bibitem [{\citenamefont {Kopnin}(2001)}]{kopnin2001theory}%
  \BibitemOpen
  \bibfield  {author} {\bibinfo {author} {\bibfnamefont {N.~B.}\ \bibnamefont
  {Kopnin}},\ }\href@noop {} {\emph {\bibinfo {title} {Vortices in Type-II
  Superconductors: Structure and Dynamics}}},\ Vol.\ \bibinfo {volume} {110}\
  (\bibinfo  {publisher} {Oxford University Press},\ \bibinfo {year}
  {2001})\BibitemShut {NoStop}%
\bibitem [{\citenamefont {Kosterlitz}\ and\ \citenamefont
  {Thouless}(1973)}]{JMKosterlitz_1973}%
  \BibitemOpen
  \bibfield  {author} {\bibinfo {author} {\bibfnamefont {J.~M.}\ \bibnamefont
  {Kosterlitz}}\ and\ \bibinfo {author} {\bibfnamefont {D.~J.}\ \bibnamefont
  {Thouless}},\ }\bibfield  {title} {\bibinfo {title} {Ordering, metastability
  and phase transitions in two-dimensional systems},\ }\href
  {https://doi.org/10.1088/0022-3719/6/7/010} {\bibfield  {journal} {\bibinfo
  {journal} {Journal of Physics C: Solid State Physics}\ }\textbf {\bibinfo
  {volume} {6}},\ \bibinfo {pages} {1181} (\bibinfo {year} {1973})}\BibitemShut
  {NoStop}%
\bibitem [{\citenamefont {Kosterlitz}(2017)}]{RevModPhys.89.040501}%
  \BibitemOpen
  \bibfield  {author} {\bibinfo {author} {\bibfnamefont {J.~M.}\ \bibnamefont
  {Kosterlitz}},\ }\bibfield  {title} {\bibinfo {title} {Nobel lecture:
  Topological defects and phase transitions},\ }\href
  {https://doi.org/10.1103/RevModPhys.89.040501} {\bibfield  {journal}
  {\bibinfo  {journal} {Rev. Mod. Phys.}\ }\textbf {\bibinfo {volume} {89}},\
  \bibinfo {pages} {040501} (\bibinfo {year} {2017})}\BibitemShut {NoStop}%
\bibitem [{\citenamefont {Abrikosov}(1957)}]{ABRIKOSOV1957199}%
  \BibitemOpen
  \bibfield  {author} {\bibinfo {author} {\bibfnamefont {A.}~\bibnamefont
  {Abrikosov}},\ }\bibfield  {title} {\bibinfo {title} {The magnetic properties
  of superconducting alloys},\ }\href
  {https://doi.org/https://doi.org/10.1016/0022-3697(57)90083-5} {\bibfield
  {journal} {\bibinfo  {journal} {Journal of Physics and Chemistry of Solids}\
  }\textbf {\bibinfo {volume} {2}},\ \bibinfo {pages} {199} (\bibinfo {year}
  {1957})}\BibitemShut {NoStop}%
\bibitem [{\citenamefont {Tinkham}(2004)}]{Tinkham}%
  \BibitemOpen
  \bibfield  {author} {\bibinfo {author} {\bibfnamefont {M.}~\bibnamefont
  {Tinkham}},\ }\href@noop {} {\emph {\bibinfo {title} {Introduction to
  Superconductivity,Dover Books on Physics Series}}}\ (\bibinfo  {publisher}
  {Dover, New York},\ \bibinfo {year} {2004})\BibitemShut {NoStop}%
\bibitem [{\citenamefont {Datta}\ \emph {et~al.}(2023)\citenamefont {Datta},
  \citenamefont {Changlani}, \citenamefont {Yang},\ and\ \citenamefont
  {Ghosal}}]{PhysRevB.107.L140505}%
  \BibitemOpen
  \bibfield  {author} {\bibinfo {author} {\bibfnamefont {A.}~\bibnamefont
  {Datta}}, \bibinfo {author} {\bibfnamefont {H.~J.}\ \bibnamefont
  {Changlani}}, \bibinfo {author} {\bibfnamefont {K.}~\bibnamefont {Yang}},\
  and\ \bibinfo {author} {\bibfnamefont {A.}~\bibnamefont {Ghosal}},\
  }\bibfield  {title} {\bibinfo {title} {Enigma of the vortex state in a
  strongly correlated $d$-wave superconductor},\ }\href
  {https://doi.org/10.1103/PhysRevB.107.L140505} {\bibfield  {journal}
  {\bibinfo  {journal} {Phys. Rev. B}\ }\textbf {\bibinfo {volume} {107}},\
  \bibinfo {pages} {L140505} (\bibinfo {year} {2023})}\BibitemShut {NoStop}%
\bibitem [{\citenamefont {Paramekanti}\ \emph {et~al.}(2001)\citenamefont
  {Paramekanti}, \citenamefont {Randeria},\ and\ \citenamefont
  {Trivedi}}]{PhysRevLett.87.217002}%
  \BibitemOpen
  \bibfield  {author} {\bibinfo {author} {\bibfnamefont {A.}~\bibnamefont
  {Paramekanti}}, \bibinfo {author} {\bibfnamefont {M.}~\bibnamefont
  {Randeria}},\ and\ \bibinfo {author} {\bibfnamefont {N.}~\bibnamefont
  {Trivedi}},\ }\bibfield  {title} {\bibinfo {title} {Projected wave functions
  and high temperature superconductivity},\ }\href
  {https://doi.org/10.1103/PhysRevLett.87.217002} {\bibfield  {journal}
  {\bibinfo  {journal} {Phys. Rev. Lett.}\ }\textbf {\bibinfo {volume} {87}},\
  \bibinfo {pages} {217002} (\bibinfo {year} {2001})}\BibitemShut {NoStop}%
\bibitem [{\citenamefont {Ioffe}\ and\ \citenamefont
  {Millis}(2002)}]{PhysRevB.66.094513}%
  \BibitemOpen
  \bibfield  {author} {\bibinfo {author} {\bibfnamefont {L.~B.}\ \bibnamefont
  {Ioffe}}\ and\ \bibinfo {author} {\bibfnamefont {A.~J.}\ \bibnamefont
  {Millis}},\ }\bibfield  {title} {\bibinfo {title} {Big fast vortices in the
  d-wave resonating valence bond theory of high-temperature
  superconductivity},\ }\href {https://doi.org/10.1103/PhysRevB.66.094513}
  {\bibfield  {journal} {\bibinfo  {journal} {Phys. Rev. B}\ }\textbf {\bibinfo
  {volume} {66}},\ \bibinfo {pages} {094513} (\bibinfo {year}
  {2002})}\BibitemShut {NoStop}%
\bibitem [{\citenamefont {Lee}\ \emph {et~al.}(2006)\citenamefont {Lee},
  \citenamefont {Nagaosa},\ and\ \citenamefont {Wen}}]{RevModPhys.78.17}%
  \BibitemOpen
  \bibfield  {author} {\bibinfo {author} {\bibfnamefont {P.~A.}\ \bibnamefont
  {Lee}}, \bibinfo {author} {\bibfnamefont {N.}~\bibnamefont {Nagaosa}},\ and\
  \bibinfo {author} {\bibfnamefont {X.-G.}\ \bibnamefont {Wen}},\ }\bibfield
  {title} {\bibinfo {title} {Doping a mott insulator: Physics of
  high-temperature superconductivity},\ }\href
  {https://doi.org/10.1103/RevModPhys.78.17} {\bibfield  {journal} {\bibinfo
  {journal} {Rev. Mod. Phys.}\ }\textbf {\bibinfo {volume} {78}},\ \bibinfo
  {pages} {17} (\bibinfo {year} {2006})}\BibitemShut {NoStop}%
\bibitem [{\citenamefont {Fischer}\ \emph {et~al.}(2007)\citenamefont
  {Fischer}, \citenamefont {Kugler}, \citenamefont {Maggio-Aprile},
  \citenamefont {Berthod},\ and\ \citenamefont {Renner}}]{RevModPhys.79.353}%
  \BibitemOpen
  \bibfield  {author} {\bibinfo {author} {\bibfnamefont {O.}~\bibnamefont
  {Fischer}}, \bibinfo {author} {\bibfnamefont {M.}~\bibnamefont {Kugler}},
  \bibinfo {author} {\bibfnamefont {I.}~\bibnamefont {Maggio-Aprile}}, \bibinfo
  {author} {\bibfnamefont {C.}~\bibnamefont {Berthod}},\ and\ \bibinfo {author}
  {\bibfnamefont {C.}~\bibnamefont {Renner}},\ }\bibfield  {title} {\bibinfo
  {title} {Scanning tunneling spectroscopy of high-temperature
  superconductors},\ }\href {https://doi.org/10.1103/RevModPhys.79.353}
  {\bibfield  {journal} {\bibinfo  {journal} {Rev. Mod. Phys.}\ }\textbf
  {\bibinfo {volume} {79}},\ \bibinfo {pages} {353} (\bibinfo {year}
  {2007})}\BibitemShut {NoStop}%
\bibitem [{\citenamefont {Wang}\ and\ \citenamefont
  {MacDonald}(1995)}]{PhysRevB.52.R3876}%
  \BibitemOpen
  \bibfield  {author} {\bibinfo {author} {\bibfnamefont {Y.}~\bibnamefont
  {Wang}}\ and\ \bibinfo {author} {\bibfnamefont {A.~H.}\ \bibnamefont
  {MacDonald}},\ }\bibfield  {title} {\bibinfo {title} {Mixed-state
  quasiparticle spectrum for d-wave superconductors},\ }\href
  {https://doi.org/10.1103/PhysRevB.52.R3876} {\bibfield  {journal} {\bibinfo
  {journal} {Phys. Rev. B}\ }\textbf {\bibinfo {volume} {52}},\ \bibinfo
  {pages} {R3876} (\bibinfo {year} {1995})}\BibitemShut {NoStop}%
\bibitem [{\citenamefont {Tsuchiura}\ \emph {et~al.}(2003)\citenamefont
  {Tsuchiura}, \citenamefont {Ogata}, \citenamefont {Tanaka},\ and\
  \citenamefont {Kashiwaya}}]{PhysRevB.68.012509}%
  \BibitemOpen
  \bibfield  {author} {\bibinfo {author} {\bibfnamefont {H.}~\bibnamefont
  {Tsuchiura}}, \bibinfo {author} {\bibfnamefont {M.}~\bibnamefont {Ogata}},
  \bibinfo {author} {\bibfnamefont {Y.}~\bibnamefont {Tanaka}},\ and\ \bibinfo
  {author} {\bibfnamefont {S.}~\bibnamefont {Kashiwaya}},\ }\bibfield  {title}
  {\bibinfo {title} {Electronic states around a vortex core in high-${T}_{c}$
  superconductors based on the $t$-$j$ model},\ }\href
  {https://doi.org/10.1103/PhysRevB.68.012509} {\bibfield  {journal} {\bibinfo
  {journal} {Phys. Rev. B}\ }\textbf {\bibinfo {volume} {68}},\ \bibinfo
  {pages} {012509} (\bibinfo {year} {2003})}\BibitemShut {NoStop}%
\bibitem [{\citenamefont {Banerjee}\ \emph {et~al.}(2022)\citenamefont
  {Banerjee}, \citenamefont {P\'epin},\ and\ \citenamefont
  {Ghosal}}]{PhysRevB.105.134505}%
  \BibitemOpen
  \bibfield  {author} {\bibinfo {author} {\bibfnamefont {A.}~\bibnamefont
  {Banerjee}}, \bibinfo {author} {\bibfnamefont {C.}~\bibnamefont {P\'epin}},\
  and\ \bibinfo {author} {\bibfnamefont {A.}~\bibnamefont {Ghosal}},\
  }\bibfield  {title} {\bibinfo {title} {Charge, bond, and pair density wave
  orders in a strongly correlated system},\ }\href
  {https://doi.org/10.1103/PhysRevB.105.134505} {\bibfield  {journal} {\bibinfo
   {journal} {Phys. Rev. B}\ }\textbf {\bibinfo {volume} {105}},\ \bibinfo
  {pages} {134505} (\bibinfo {year} {2022})}\BibitemShut {NoStop}%
\bibitem [{\citenamefont {Dai}\ \emph {et~al.}(2018)\citenamefont {Dai},
  \citenamefont {Zhang}, \citenamefont {Senthil},\ and\ \citenamefont
  {Lee}}]{PhysRevB.97.174511}%
  \BibitemOpen
  \bibfield  {author} {\bibinfo {author} {\bibfnamefont {Z.}~\bibnamefont
  {Dai}}, \bibinfo {author} {\bibfnamefont {Y.-H.}\ \bibnamefont {Zhang}},
  \bibinfo {author} {\bibfnamefont {T.}~\bibnamefont {Senthil}},\ and\ \bibinfo
  {author} {\bibfnamefont {P.~A.}\ \bibnamefont {Lee}},\ }\bibfield  {title}
  {\bibinfo {title} {Pair-density waves, charge-density waves, and vortices in
  high-${T}_{c}$ cuprates},\ }\href
  {https://doi.org/10.1103/PhysRevB.97.174511} {\bibfield  {journal} {\bibinfo
  {journal} {Phys. Rev. B}\ }\textbf {\bibinfo {volume} {97}},\ \bibinfo
  {pages} {174511} (\bibinfo {year} {2018})}\BibitemShut {NoStop}%
\bibitem [{\citenamefont {Liu}\ \emph {et~al.}(2023)\citenamefont {Liu},
  \citenamefont {Tu}, \citenamefont {Chern},\ and\ \citenamefont
  {Lee}}]{PhysRevResearch.5.033028}%
  \BibitemOpen
  \bibfield  {author} {\bibinfo {author} {\bibfnamefont {Y.-H.}\ \bibnamefont
  {Liu}}, \bibinfo {author} {\bibfnamefont {W.-L.}\ \bibnamefont {Tu}},
  \bibinfo {author} {\bibfnamefont {G.-W.}\ \bibnamefont {Chern}},\ and\
  \bibinfo {author} {\bibfnamefont {T.-K.}\ \bibnamefont {Lee}},\ }\bibfield
  {title} {\bibinfo {title} {Intertwined orders and electronic structure in
  superconducting vortex halos},\ }\href
  {https://doi.org/10.1103/PhysRevResearch.5.033028} {\bibfield  {journal}
  {\bibinfo  {journal} {Phys. Rev. Res.}\ }\textbf {\bibinfo {volume} {5}},\
  \bibinfo {pages} {033028} (\bibinfo {year} {2023})}\BibitemShut {NoStop}%
\bibitem [{\citenamefont {Fradkin}\ \emph {et~al.}(2015)\citenamefont
  {Fradkin}, \citenamefont {Kivelson},\ and\ \citenamefont
  {Tranquada}}]{RevModPhys.87.457}%
  \BibitemOpen
  \bibfield  {author} {\bibinfo {author} {\bibfnamefont {E.}~\bibnamefont
  {Fradkin}}, \bibinfo {author} {\bibfnamefont {S.~A.}\ \bibnamefont
  {Kivelson}},\ and\ \bibinfo {author} {\bibfnamefont {J.~M.}\ \bibnamefont
  {Tranquada}},\ }\bibfield  {title} {\bibinfo {title} {Colloquium: Theory of
  intertwined orders in high temperature superconductors},\ }\href
  {https://doi.org/10.1103/RevModPhys.87.457} {\bibfield  {journal} {\bibinfo
  {journal} {Rev. Mod. Phys.}\ }\textbf {\bibinfo {volume} {87}},\ \bibinfo
  {pages} {457} (\bibinfo {year} {2015})}\BibitemShut {NoStop}%
\bibitem [{\citenamefont {Dagotto}(2005)}]{doi:10.1126/science.1107559}%
  \BibitemOpen
  \bibfield  {author} {\bibinfo {author} {\bibfnamefont {E.}~\bibnamefont
  {Dagotto}},\ }\bibfield  {title} {\bibinfo {title} {Complexity in strongly
  correlated electronic systems},\ }\href
  {https://doi.org/10.1126/science.1107559} {\bibfield  {journal} {\bibinfo
  {journal} {Science}\ }\textbf {\bibinfo {volume} {309}},\ \bibinfo {pages}
  {257} (\bibinfo {year} {2005})}\BibitemShut {NoStop}%
\bibitem [{\citenamefont {Kivelson}\ \emph {et~al.}(2003)\citenamefont
  {Kivelson}, \citenamefont {Bindloss}, \citenamefont {Fradkin}, \citenamefont
  {Oganesyan}, \citenamefont {Tranquada}, \citenamefont {Kapitulnik},\ and\
  \citenamefont {Howald}}]{RevModPhys.75.1201}%
  \BibitemOpen
  \bibfield  {author} {\bibinfo {author} {\bibfnamefont {S.~A.}\ \bibnamefont
  {Kivelson}}, \bibinfo {author} {\bibfnamefont {I.~P.}\ \bibnamefont
  {Bindloss}}, \bibinfo {author} {\bibfnamefont {E.}~\bibnamefont {Fradkin}},
  \bibinfo {author} {\bibfnamefont {V.}~\bibnamefont {Oganesyan}}, \bibinfo
  {author} {\bibfnamefont {J.~M.}\ \bibnamefont {Tranquada}}, \bibinfo {author}
  {\bibfnamefont {A.}~\bibnamefont {Kapitulnik}},\ and\ \bibinfo {author}
  {\bibfnamefont {C.}~\bibnamefont {Howald}},\ }\bibfield  {title} {\bibinfo
  {title} {How to detect fluctuating stripes in the high-temperature
  superconductors},\ }\href {https://doi.org/10.1103/RevModPhys.75.1201}
  {\bibfield  {journal} {\bibinfo  {journal} {Rev. Mod. Phys.}\ }\textbf
  {\bibinfo {volume} {75}},\ \bibinfo {pages} {1201} (\bibinfo {year}
  {2003})}\BibitemShut {NoStop}%
\bibitem [{\citenamefont {Comin}\ and\ \citenamefont
  {Damascelli}(2016)}]{doi:10.1146/annurev-conmatphys-031115-011401}%
  \BibitemOpen
  \bibfield  {author} {\bibinfo {author} {\bibfnamefont {R.}~\bibnamefont
  {Comin}}\ and\ \bibinfo {author} {\bibfnamefont {A.}~\bibnamefont
  {Damascelli}},\ }\bibfield  {title} {\bibinfo {title} {Resonant x-ray
  scattering studies of charge order in cuprates},\ }\href
  {https://doi.org/10.1146/annurev-conmatphys-031115-011401} {\bibfield
  {journal} {\bibinfo  {journal} {Annual Review of Condensed Matter Physics}\
  }\textbf {\bibinfo {volume} {7}},\ \bibinfo {pages} {369} (\bibinfo {year}
  {2016})}\BibitemShut {NoStop}%
\bibitem [{\citenamefont {Ghiringhelli}\ \emph {et~al.}(2012)\citenamefont
  {Ghiringhelli}, \citenamefont {Le~Tacon}, \citenamefont {Minola},
  \citenamefont {Blanco-Canosa}, \citenamefont {Mazzoli}, \citenamefont
  {Brookes}, \citenamefont {De~Luca}, \citenamefont {Frano}, \citenamefont
  {Hawthorn}, \citenamefont {He} \emph {et~al.}}]{doi:10.1126/science.1223532}%
  \BibitemOpen
  \bibfield  {author} {\bibinfo {author} {\bibfnamefont {G.}~\bibnamefont
  {Ghiringhelli}}, \bibinfo {author} {\bibfnamefont {M.}~\bibnamefont
  {Le~Tacon}}, \bibinfo {author} {\bibfnamefont {M.}~\bibnamefont {Minola}},
  \bibinfo {author} {\bibfnamefont {S.}~\bibnamefont {Blanco-Canosa}}, \bibinfo
  {author} {\bibfnamefont {C.}~\bibnamefont {Mazzoli}}, \bibinfo {author}
  {\bibfnamefont {N.}~\bibnamefont {Brookes}}, \bibinfo {author} {\bibfnamefont
  {G.}~\bibnamefont {De~Luca}}, \bibinfo {author} {\bibfnamefont
  {A.}~\bibnamefont {Frano}}, \bibinfo {author} {\bibfnamefont
  {D.}~\bibnamefont {Hawthorn}}, \bibinfo {author} {\bibfnamefont
  {F.}~\bibnamefont {He}}, \emph {et~al.},\ }\bibfield  {title} {\bibinfo
  {title} {Long-range incommensurate charge fluctuations in (\text{Y},
  \text{Nd})$\text{Ba}_2\text{Cu}_3\text{O}_{6+x}$},\ }\href
  {https://www.science.org/doi/abs/10.1126/science.1223532} {\bibfield
  {journal} {\bibinfo  {journal} {Science}\ }\textbf {\bibinfo {volume}
  {337}},\ \bibinfo {pages} {821} (\bibinfo {year} {2012})}\BibitemShut
  {NoStop}%
\bibitem [{\citenamefont {Blackburn}\ \emph {et~al.}(2013)\citenamefont
  {Blackburn}, \citenamefont {Chang}, \citenamefont {H\"ucker}, \citenamefont
  {Holmes}, \citenamefont {Christensen}, \citenamefont {Liang}, \citenamefont
  {Bonn}, \citenamefont {Hardy}, \citenamefont {R\"utt}, \citenamefont
  {Gutowski}, \citenamefont {Zimmermann}, \citenamefont {Forgan},\ and\
  \citenamefont {Hayden}}]{PhysRevLett.110.137004}%
  \BibitemOpen
  \bibfield  {author} {\bibinfo {author} {\bibfnamefont {E.}~\bibnamefont
  {Blackburn}}, \bibinfo {author} {\bibfnamefont {J.}~\bibnamefont {Chang}},
  \bibinfo {author} {\bibfnamefont {M.}~\bibnamefont {H\"ucker}}, \bibinfo
  {author} {\bibfnamefont {A.~T.}\ \bibnamefont {Holmes}}, \bibinfo {author}
  {\bibfnamefont {N.~B.}\ \bibnamefont {Christensen}}, \bibinfo {author}
  {\bibfnamefont {R.}~\bibnamefont {Liang}}, \bibinfo {author} {\bibfnamefont
  {D.~A.}\ \bibnamefont {Bonn}}, \bibinfo {author} {\bibfnamefont {W.~N.}\
  \bibnamefont {Hardy}}, \bibinfo {author} {\bibfnamefont {U.}~\bibnamefont
  {R\"utt}}, \bibinfo {author} {\bibfnamefont {O.}~\bibnamefont {Gutowski}},
  \bibinfo {author} {\bibfnamefont {M.~v.}\ \bibnamefont {Zimmermann}},
  \bibinfo {author} {\bibfnamefont {E.~M.}\ \bibnamefont {Forgan}},\ and\
  \bibinfo {author} {\bibfnamefont {S.~M.}\ \bibnamefont {Hayden}},\ }\bibfield
   {title} {\bibinfo {title} {X-ray diffraction observations of a
  charge-density-wave order in superconducting ortho-ii
  $\text{YBa}_2\text{Cu}_3\text{O}_{6.54}$ single crystals in zero magnetic
  field},\ }\href {https://doi.org/10.1103/PhysRevLett.110.137004} {\bibfield
  {journal} {\bibinfo  {journal} {Phys. Rev. Lett.}\ }\textbf {\bibinfo
  {volume} {110}},\ \bibinfo {pages} {137004} (\bibinfo {year}
  {2013})}\BibitemShut {NoStop}%
\bibitem [{\citenamefont {Blanco-Canosa}\ \emph {et~al.}(2014)\citenamefont
  {Blanco-Canosa}, \citenamefont {Frano}, \citenamefont {Schierle},
  \citenamefont {Porras}, \citenamefont {Loew}, \citenamefont {Minola},
  \citenamefont {Bluschke}, \citenamefont {Weschke}, \citenamefont {Keimer},\
  and\ \citenamefont {Le~Tacon}}]{PhysRevB.90.054513}%
  \BibitemOpen
  \bibfield  {author} {\bibinfo {author} {\bibfnamefont {S.}~\bibnamefont
  {Blanco-Canosa}}, \bibinfo {author} {\bibfnamefont {A.}~\bibnamefont
  {Frano}}, \bibinfo {author} {\bibfnamefont {E.}~\bibnamefont {Schierle}},
  \bibinfo {author} {\bibfnamefont {J.}~\bibnamefont {Porras}}, \bibinfo
  {author} {\bibfnamefont {T.}~\bibnamefont {Loew}}, \bibinfo {author}
  {\bibfnamefont {M.}~\bibnamefont {Minola}}, \bibinfo {author} {\bibfnamefont
  {M.}~\bibnamefont {Bluschke}}, \bibinfo {author} {\bibfnamefont
  {E.}~\bibnamefont {Weschke}}, \bibinfo {author} {\bibfnamefont
  {B.}~\bibnamefont {Keimer}},\ and\ \bibinfo {author} {\bibfnamefont
  {M.}~\bibnamefont {Le~Tacon}},\ }\bibfield  {title} {\bibinfo {title}
  {Resonant x-ray scattering study of charge-density wave correlations in
  $\text{YBa}_2\text{Cu}_3\text{O}_{6 + x}$},\ }\href
  {https://doi.org/10.1103/PhysRevB.90.054513} {\bibfield  {journal} {\bibinfo
  {journal} {Phys. Rev. B}\ }\textbf {\bibinfo {volume} {90}},\ \bibinfo
  {pages} {054513} (\bibinfo {year} {2014})}\BibitemShut {NoStop}%
\bibitem [{\citenamefont {Wu}\ \emph {et~al.}(2011)\citenamefont {Wu},
  \citenamefont {Mayaffre}, \citenamefont {Kr{\"a}mer}, \citenamefont
  {Horvati{\'c}}, \citenamefont {Berthier}, \citenamefont {Hardy},
  \citenamefont {Liang}, \citenamefont {Bonn},\ and\ \citenamefont
  {Julien}}]{wu2011magnetic}%
  \BibitemOpen
  \bibfield  {author} {\bibinfo {author} {\bibfnamefont {T.}~\bibnamefont
  {Wu}}, \bibinfo {author} {\bibfnamefont {H.}~\bibnamefont {Mayaffre}},
  \bibinfo {author} {\bibfnamefont {S.}~\bibnamefont {Kr{\"a}mer}}, \bibinfo
  {author} {\bibfnamefont {M.}~\bibnamefont {Horvati{\'c}}}, \bibinfo {author}
  {\bibfnamefont {C.}~\bibnamefont {Berthier}}, \bibinfo {author}
  {\bibfnamefont {W.}~\bibnamefont {Hardy}}, \bibinfo {author} {\bibfnamefont
  {R.}~\bibnamefont {Liang}}, \bibinfo {author} {\bibfnamefont
  {D.}~\bibnamefont {Bonn}},\ and\ \bibinfo {author} {\bibfnamefont {M.-H.}\
  \bibnamefont {Julien}},\ }\bibfield  {title} {\bibinfo {title}
  {Magnetic-field-induced charge-stripe order in the high-temperature
  superconductor $\text{YBa}_2\text{Cu}_3\text{O}_{y}$},\ }\href
  {https://doi.org/10.1038/nature10345} {\bibfield  {journal} {\bibinfo
  {journal} {Nature}\ }\textbf {\bibinfo {volume} {477}},\ \bibinfo {pages}
  {191} (\bibinfo {year} {2011})}\BibitemShut {NoStop}%
\bibitem [{\citenamefont {Hamidian}\ \emph
  {et~al.}(2016{\natexlab{a}})\citenamefont {Hamidian}, \citenamefont {Edkins},
  \citenamefont {Kim}, \citenamefont {Davis}, \citenamefont {Mackenzie},
  \citenamefont {Eisaki}, \citenamefont {Uchida}, \citenamefont {Lawler},
  \citenamefont {Kim}, \citenamefont {Sachdev} \emph
  {et~al.}}]{hamidian2016atomic}%
  \BibitemOpen
  \bibfield  {author} {\bibinfo {author} {\bibfnamefont {M.}~\bibnamefont
  {Hamidian}}, \bibinfo {author} {\bibfnamefont {S.~D.}\ \bibnamefont
  {Edkins}}, \bibinfo {author} {\bibfnamefont {C.~K.}\ \bibnamefont {Kim}},
  \bibinfo {author} {\bibfnamefont {J.~C.}\ \bibnamefont {Davis}}, \bibinfo
  {author} {\bibfnamefont {A.}~\bibnamefont {Mackenzie}}, \bibinfo {author}
  {\bibfnamefont {H.}~\bibnamefont {Eisaki}}, \bibinfo {author} {\bibfnamefont
  {S.}~\bibnamefont {Uchida}}, \bibinfo {author} {\bibfnamefont
  {M.}~\bibnamefont {Lawler}}, \bibinfo {author} {\bibfnamefont {E.-A.}\
  \bibnamefont {Kim}}, \bibinfo {author} {\bibfnamefont {S.}~\bibnamefont
  {Sachdev}}, \emph {et~al.},\ }\bibfield  {title} {\bibinfo {title}
  {Atomic-scale electronic structure of the cuprate d-symmetry form factor
  density wave state},\ }\href {https://doi.org/10.1038/nphys3519} {\bibfield
  {journal} {\bibinfo  {journal} {Nature Physics}\ }\textbf {\bibinfo {volume}
  {12}},\ \bibinfo {pages} {150} (\bibinfo {year}
  {2016}{\natexlab{a}})}\BibitemShut {NoStop}%
\bibitem [{\citenamefont {Zhou}\ \emph {et~al.}(2017)\citenamefont {Zhou},
  \citenamefont {Hirata}, \citenamefont {Wu}, \citenamefont {Vinograd},
  \citenamefont {Mayaffre}, \citenamefont {Kr{\"a}mer}, \citenamefont {Reyes},
  \citenamefont {Kuhns}, \citenamefont {Liang}, \citenamefont {Hardy} \emph
  {et~al.}}]{zhou2017spin}%
  \BibitemOpen
  \bibfield  {author} {\bibinfo {author} {\bibfnamefont {R.}~\bibnamefont
  {Zhou}}, \bibinfo {author} {\bibfnamefont {M.}~\bibnamefont {Hirata}},
  \bibinfo {author} {\bibfnamefont {T.}~\bibnamefont {Wu}}, \bibinfo {author}
  {\bibfnamefont {I.}~\bibnamefont {Vinograd}}, \bibinfo {author}
  {\bibfnamefont {H.}~\bibnamefont {Mayaffre}}, \bibinfo {author}
  {\bibfnamefont {S.}~\bibnamefont {Kr{\"a}mer}}, \bibinfo {author}
  {\bibfnamefont {A.~P.}\ \bibnamefont {Reyes}}, \bibinfo {author}
  {\bibfnamefont {P.~L.}\ \bibnamefont {Kuhns}}, \bibinfo {author}
  {\bibfnamefont {R.}~\bibnamefont {Liang}}, \bibinfo {author} {\bibfnamefont
  {W.}~\bibnamefont {Hardy}}, \emph {et~al.},\ }\bibfield  {title} {\bibinfo
  {title} {Spin susceptibility of charge-ordered yba2cu3o y across the upper
  critical field},\ }\href
  {https://www.pnas.org/doi/abs/10.1073/pnas.1711445114} {\bibfield  {journal}
  {\bibinfo  {journal} {Proceedings of the National Academy of Sciences}\
  }\textbf {\bibinfo {volume} {114}},\ \bibinfo {pages} {13148} (\bibinfo
  {year} {2017})}\BibitemShut {NoStop}%
\bibitem [{\citenamefont {Wen}\ \emph {et~al.}(2023)\citenamefont {Wen},
  \citenamefont {He}, \citenamefont {Jang}, \citenamefont {Nojiri},
  \citenamefont {Matsuzawa}, \citenamefont {Song}, \citenamefont {Chollet},
  \citenamefont {Zhu}, \citenamefont {Liu}, \citenamefont {Fujita} \emph
  {et~al.}}]{wen2023enhanced}%
  \BibitemOpen
  \bibfield  {author} {\bibinfo {author} {\bibfnamefont {J.-J.}\ \bibnamefont
  {Wen}}, \bibinfo {author} {\bibfnamefont {W.}~\bibnamefont {He}}, \bibinfo
  {author} {\bibfnamefont {H.}~\bibnamefont {Jang}}, \bibinfo {author}
  {\bibfnamefont {H.}~\bibnamefont {Nojiri}}, \bibinfo {author} {\bibfnamefont
  {S.}~\bibnamefont {Matsuzawa}}, \bibinfo {author} {\bibfnamefont
  {S.}~\bibnamefont {Song}}, \bibinfo {author} {\bibfnamefont {M.}~\bibnamefont
  {Chollet}}, \bibinfo {author} {\bibfnamefont {D.}~\bibnamefont {Zhu}},
  \bibinfo {author} {\bibfnamefont {Y.-J.}\ \bibnamefont {Liu}}, \bibinfo
  {author} {\bibfnamefont {M.}~\bibnamefont {Fujita}}, \emph {et~al.},\
  }\bibfield  {title} {\bibinfo {title} {Enhanced charge density wave with
  mobile superconducting vortices in $\text{La}_{1.885}\text{Sr}_{0.
  115}\text{Cu}\text{O}_{4}$},\ }\href
  {https://doi.org/10.1038/s41467-023-36203-x} {\bibfield  {journal} {\bibinfo
  {journal} {Nature communications}\ }\textbf {\bibinfo {volume} {14}},\
  \bibinfo {pages} {733} (\bibinfo {year} {2023})}\BibitemShut {NoStop}%
\bibitem [{\citenamefont {Hoffman}\ \emph {et~al.}(2002)\citenamefont
  {Hoffman}, \citenamefont {Hudson}, \citenamefont {Lang}, \citenamefont
  {Madhavan}, \citenamefont {Eisaki}, \citenamefont {Uchida},\ and\
  \citenamefont {Davis}}]{hoffman2002}%
  \BibitemOpen
  \bibfield  {author} {\bibinfo {author} {\bibfnamefont {J.~E.}\ \bibnamefont
  {Hoffman}}, \bibinfo {author} {\bibfnamefont {E.~W.}\ \bibnamefont {Hudson}},
  \bibinfo {author} {\bibfnamefont {K.~M.}\ \bibnamefont {Lang}}, \bibinfo
  {author} {\bibfnamefont {V.}~\bibnamefont {Madhavan}}, \bibinfo {author}
  {\bibfnamefont {H.}~\bibnamefont {Eisaki}}, \bibinfo {author} {\bibfnamefont
  {S.}~\bibnamefont {Uchida}},\ and\ \bibinfo {author} {\bibfnamefont {J.~C.}\
  \bibnamefont {Davis}},\ }\bibfield  {title} {\bibinfo {title} {A four unit
  cell periodic pattern of quasi-particle states surrounding vortex cores in
  {Bi\textsubscript{2}Sr\textsubscript{2}CaCu\textsubscript{2}O\textsubscript{8+\ensuremath{\delta}}}},\
  }\href {https://doi.org/10.1126/science.1066974} {\bibfield  {journal}
  {\bibinfo  {journal} {Science}\ }\textbf {\bibinfo {volume} {295}},\ \bibinfo
  {pages} {466} (\bibinfo {year} {2002})}\BibitemShut {NoStop}%
\bibitem [{\citenamefont {Wu}\ \emph {et~al.}(2013)\citenamefont {Wu},
  \citenamefont {Mayaffre}, \citenamefont {Kr{\"a}mer}, \citenamefont
  {Horvati{\'{c}}}, \citenamefont {Berthier}, \citenamefont {Kuhns},
  \citenamefont {Reyes}, \citenamefont {Liang}, \citenamefont {Hardy},
  \citenamefont {Bonn},\ and\ \citenamefont {Julien}}]{Wu2013}%
  \BibitemOpen
  \bibfield  {author} {\bibinfo {author} {\bibfnamefont {T.}~\bibnamefont
  {Wu}}, \bibinfo {author} {\bibfnamefont {H.}~\bibnamefont {Mayaffre}},
  \bibinfo {author} {\bibfnamefont {S.}~\bibnamefont {Kr{\"a}mer}}, \bibinfo
  {author} {\bibfnamefont {M.}~\bibnamefont {Horvati{\'{c}}}}, \bibinfo
  {author} {\bibfnamefont {C.}~\bibnamefont {Berthier}}, \bibinfo {author}
  {\bibfnamefont {P.~L.}\ \bibnamefont {Kuhns}}, \bibinfo {author}
  {\bibfnamefont {A.~P.}\ \bibnamefont {Reyes}}, \bibinfo {author}
  {\bibfnamefont {R.}~\bibnamefont {Liang}}, \bibinfo {author} {\bibfnamefont
  {W.~N.}\ \bibnamefont {Hardy}}, \bibinfo {author} {\bibfnamefont {D.~A.}\
  \bibnamefont {Bonn}},\ and\ \bibinfo {author} {\bibfnamefont {M.-H.}\
  \bibnamefont {Julien}},\ }\bibfield  {title} {\bibinfo {title} {Emergence of
  charge order from the vortex state of a high-temperature superconductor},\
  }\href {https://doi.org/10.1038/ncomms3113} {\bibfield  {journal} {\bibinfo
  {journal} {Nature Communications}\ }\textbf {\bibinfo {volume} {4}},\
  \bibinfo {pages} {2113} (\bibinfo {year} {2013})}\BibitemShut {NoStop}%
\bibitem [{\citenamefont {Tu}\ and\ \citenamefont {Lee}(2016)}]{Tu2016}%
  \BibitemOpen
  \bibfield  {author} {\bibinfo {author} {\bibfnamefont {W.-L.}\ \bibnamefont
  {Tu}}\ and\ \bibinfo {author} {\bibfnamefont {T.-K.}\ \bibnamefont {Lee}},\
  }\bibfield  {title} {\bibinfo {title} {Genesis of charge orders in high
  temperature superconductors},\ }\href {https://doi.org/10.1038/srep18675}
  {\bibfield  {journal} {\bibinfo  {journal} {Scientific Reports}\ }\textbf
  {\bibinfo {volume} {6}},\ \bibinfo {pages} {18675} (\bibinfo {year}
  {2016})}\BibitemShut {NoStop}%
\bibitem [{\citenamefont {Edkins}\ \emph {et~al.}(2019)\citenamefont {Edkins},
  \citenamefont {Kostin}, \citenamefont {Fujita}, \citenamefont {Mackenzie},
  \citenamefont {Eisaki}, \citenamefont {Uchida}, \citenamefont {Sachdev},
  \citenamefont {Lawler}, \citenamefont {Kim}, \citenamefont {Davis},\ and\
  \citenamefont {Hamidian}}]{doi:10.1126/science.aat1773}%
  \BibitemOpen
  \bibfield  {author} {\bibinfo {author} {\bibfnamefont {S.~D.}\ \bibnamefont
  {Edkins}}, \bibinfo {author} {\bibfnamefont {A.}~\bibnamefont {Kostin}},
  \bibinfo {author} {\bibfnamefont {K.}~\bibnamefont {Fujita}}, \bibinfo
  {author} {\bibfnamefont {A.~P.}\ \bibnamefont {Mackenzie}}, \bibinfo {author}
  {\bibfnamefont {H.}~\bibnamefont {Eisaki}}, \bibinfo {author} {\bibfnamefont
  {S.}~\bibnamefont {Uchida}}, \bibinfo {author} {\bibfnamefont
  {S.}~\bibnamefont {Sachdev}}, \bibinfo {author} {\bibfnamefont {M.~J.}\
  \bibnamefont {Lawler}}, \bibinfo {author} {\bibfnamefont {E.-A.}\
  \bibnamefont {Kim}}, \bibinfo {author} {\bibfnamefont {J.~C.~S.}\
  \bibnamefont {Davis}},\ and\ \bibinfo {author} {\bibfnamefont {M.~H.}\
  \bibnamefont {Hamidian}},\ }\bibfield  {title} {\bibinfo {title} {Magnetic
  field–induced pair density wave state in the cuprate vortex halo},\ }\href
  {https://doi.org/10.1126/science.aat1773} {\bibfield  {journal} {\bibinfo
  {journal} {Science}\ }\textbf {\bibinfo {volume} {364}},\ \bibinfo {pages}
  {976} (\bibinfo {year} {2019})}\BibitemShut {NoStop}%
\bibitem [{\citenamefont {Zou}\ \emph {et~al.}(2024)\citenamefont {Zou},
  \citenamefont {Choi}, \citenamefont {Li}, \citenamefont {Ye}, \citenamefont
  {Yin}, \citenamefont {Garcia-Fernandez}, \citenamefont {Agrestini},
  \citenamefont {Qiu}, \citenamefont {Cai}, \citenamefont {Xiao}, \citenamefont
  {Zhou}, \citenamefont {Zhou}, \citenamefont {Wang},\ and\ \citenamefont
  {Peng}}]{Zou2024}%
  \BibitemOpen
  \bibfield  {author} {\bibinfo {author} {\bibfnamefont {C.}~\bibnamefont
  {Zou}}, \bibinfo {author} {\bibfnamefont {J.}~\bibnamefont {Choi}}, \bibinfo
  {author} {\bibfnamefont {Q.}~\bibnamefont {Li}}, \bibinfo {author}
  {\bibfnamefont {S.}~\bibnamefont {Ye}}, \bibinfo {author} {\bibfnamefont
  {C.}~\bibnamefont {Yin}}, \bibinfo {author} {\bibfnamefont {M.}~\bibnamefont
  {Garcia-Fernandez}}, \bibinfo {author} {\bibfnamefont {S.}~\bibnamefont
  {Agrestini}}, \bibinfo {author} {\bibfnamefont {Q.}~\bibnamefont {Qiu}},
  \bibinfo {author} {\bibfnamefont {X.}~\bibnamefont {Cai}}, \bibinfo {author}
  {\bibfnamefont {Q.}~\bibnamefont {Xiao}}, \bibinfo {author} {\bibfnamefont
  {X.}~\bibnamefont {Zhou}}, \bibinfo {author} {\bibfnamefont {K.-J.}\
  \bibnamefont {Zhou}}, \bibinfo {author} {\bibfnamefont {Y.}~\bibnamefont
  {Wang}},\ and\ \bibinfo {author} {\bibfnamefont {Y.}~\bibnamefont {Peng}},\
  }\bibfield  {title} {\bibinfo {title} {Evolution from a charge-ordered
  insulator to a high-temperature superconductor in
  $\text{Bi}_{2}\text{Sr}_{2}(\text{Ca},\text{Dy})\text{Cu}_{2}\text{O}_{8+\delta}$},\
  }\href {https://doi.org/10.1038/s41467-024-52124-9} {\bibfield  {journal}
  {\bibinfo  {journal} {Nature Communications}\ }\textbf {\bibinfo {volume}
  {15}},\ \bibinfo {pages} {7739} (\bibinfo {year} {2024})}\BibitemShut
  {NoStop}%
\bibitem [{\citenamefont {Wang}\ \emph {et~al.}(2018)\citenamefont {Wang},
  \citenamefont {Edkins}, \citenamefont {Hamidian}, \citenamefont {Davis},
  \citenamefont {Fradkin},\ and\ \citenamefont
  {Kivelson}}]{PhysRevB.97.174510}%
  \BibitemOpen
  \bibfield  {author} {\bibinfo {author} {\bibfnamefont {Y.}~\bibnamefont
  {Wang}}, \bibinfo {author} {\bibfnamefont {S.~D.}\ \bibnamefont {Edkins}},
  \bibinfo {author} {\bibfnamefont {M.~H.}\ \bibnamefont {Hamidian}}, \bibinfo
  {author} {\bibfnamefont {J.~C.~S.}\ \bibnamefont {Davis}}, \bibinfo {author}
  {\bibfnamefont {E.}~\bibnamefont {Fradkin}},\ and\ \bibinfo {author}
  {\bibfnamefont {S.~A.}\ \bibnamefont {Kivelson}},\ }\bibfield  {title}
  {\bibinfo {title} {Pair density waves in superconducting vortex halos},\
  }\href {https://doi.org/10.1103/PhysRevB.97.174510} {\bibfield  {journal}
  {\bibinfo  {journal} {Phys. Rev. B}\ }\textbf {\bibinfo {volume} {97}},\
  \bibinfo {pages} {174510} (\bibinfo {year} {2018})}\BibitemShut {NoStop}%
\bibitem [{\citenamefont {Franz}\ and\ \citenamefont {Te\ifmmode \check{s}\else
  \v{s}\fi{}anovi\ifmmode~\acute{c}\else
  \'{c}\fi{}}(1998)}]{PhysRevLett.80.4763}%
  \BibitemOpen
  \bibfield  {author} {\bibinfo {author} {\bibfnamefont {M.}~\bibnamefont
  {Franz}}\ and\ \bibinfo {author} {\bibfnamefont {Z.}~\bibnamefont {Te\ifmmode
  \check{s}\else \v{s}\fi{}anovi\ifmmode~\acute{c}\else \'{c}\fi{}}},\
  }\bibfield  {title} {\bibinfo {title} {Self-consistent electronic structure
  of a ${\mathit{d}}_{{\mathit{x}}^{2}\ensuremath{-}{\mathit{y}}^{2}}$ and a
  ${\mathit{d}}_{{\mathit{x}}^{2}\ensuremath{-}{\mathit{y}}^{2}}+{\mathit{id}}_{\mathit{xy}}$
  vortex},\ }\href {https://doi.org/10.1103/PhysRevLett.80.4763} {\bibfield
  {journal} {\bibinfo  {journal} {Phys. Rev. Lett.}\ }\textbf {\bibinfo
  {volume} {80}},\ \bibinfo {pages} {4763} (\bibinfo {year}
  {1998})}\BibitemShut {NoStop}%
\bibitem [{\citenamefont {Chao}\ \emph {et~al.}(1977)\citenamefont {Chao},
  \citenamefont {Spalek},\ and\ \citenamefont {Oles}}]{K_A_Chao_1977}%
  \BibitemOpen
  \bibfield  {author} {\bibinfo {author} {\bibfnamefont {K.~A.}\ \bibnamefont
  {Chao}}, \bibinfo {author} {\bibfnamefont {J.}~\bibnamefont {Spalek}},\ and\
  \bibinfo {author} {\bibfnamefont {A.~M.}\ \bibnamefont {Oles}},\ }\bibfield
  {title} {\bibinfo {title} {Kinetic exchange interaction in a narrow s-band},\
  }\href {https://doi.org/10.1088/0022-3719/10/10/002} {\bibfield  {journal}
  {\bibinfo  {journal} {Journal of Physics C: Solid State Physics}\ }\textbf
  {\bibinfo {volume} {10}},\ \bibinfo {pages} {L271} (\bibinfo {year}
  {1977})}\BibitemShut {NoStop}%
\bibitem [{\citenamefont {Anderson}\ \emph {et~al.}(2004)\citenamefont
  {Anderson}, \citenamefont {Lee}, \citenamefont {Randeria}, \citenamefont
  {Rice}, \citenamefont {Trivedi},\ and\ \citenamefont
  {Zhang}}]{PWAnderson_2004}%
  \BibitemOpen
  \bibfield  {author} {\bibinfo {author} {\bibfnamefont {P.~W.}\ \bibnamefont
  {Anderson}}, \bibinfo {author} {\bibfnamefont {P.~A.}\ \bibnamefont {Lee}},
  \bibinfo {author} {\bibfnamefont {M.}~\bibnamefont {Randeria}}, \bibinfo
  {author} {\bibfnamefont {T.~M.}\ \bibnamefont {Rice}}, \bibinfo {author}
  {\bibfnamefont {N.}~\bibnamefont {Trivedi}},\ and\ \bibinfo {author}
  {\bibfnamefont {F.~C.}\ \bibnamefont {Zhang}},\ }\bibfield  {title} {\bibinfo
  {title} {The physics behind high-temperature superconducting cuprates: the
  ‘plain vanilla’ version of rvb},\ }\href
  {https://doi.org/10.1088/0953-8984/16/24/R02} {\bibfield  {journal} {\bibinfo
   {journal} {Journal of Physics: Condensed Matter}\ }\textbf {\bibinfo
  {volume} {16}},\ \bibinfo {pages} {R755} (\bibinfo {year}
  {2004})}\BibitemShut {NoStop}%
\bibitem [{\citenamefont {Norman}\ \emph {et~al.}(1995)\citenamefont {Norman},
  \citenamefont {Randeria}, \citenamefont {Ding},\ and\ \citenamefont
  {Campuzano}}]{PhysRevB.52.615}%
  \BibitemOpen
  \bibfield  {author} {\bibinfo {author} {\bibfnamefont {M.~R.}\ \bibnamefont
  {Norman}}, \bibinfo {author} {\bibfnamefont {M.}~\bibnamefont {Randeria}},
  \bibinfo {author} {\bibfnamefont {H.}~\bibnamefont {Ding}},\ and\ \bibinfo
  {author} {\bibfnamefont {J.~C.}\ \bibnamefont {Campuzano}},\ }\bibfield
  {title} {\bibinfo {title} {Phenomenological models for the gap anisotropy of
  ${\mathrm{bi}}_{2}$${\mathrm{sr}}_{2}$${\mathrm{cacu}}_{2}$${\mathrm{o}}_{8}$
  as measured by angle-resolved photoemission spectroscopy},\ }\href
  {https://doi.org/10.1103/PhysRevB.52.615} {\bibfield  {journal} {\bibinfo
  {journal} {Phys. Rev. B}\ }\textbf {\bibinfo {volume} {52}},\ \bibinfo
  {pages} {615} (\bibinfo {year} {1995})}\BibitemShut {NoStop}%
\bibitem [{\citenamefont {Ko}\ \emph {et~al.}(2007)\citenamefont {Ko},
  \citenamefont {Nave},\ and\ \citenamefont {Lee}}]{PhysRevB.76.245113}%
  \BibitemOpen
  \bibfield  {author} {\bibinfo {author} {\bibfnamefont {W.-H.}\ \bibnamefont
  {Ko}}, \bibinfo {author} {\bibfnamefont {C.~P.}\ \bibnamefont {Nave}},\ and\
  \bibinfo {author} {\bibfnamefont {P.~A.}\ \bibnamefont {Lee}},\ }\bibfield
  {title} {\bibinfo {title} {Extended gutzwiller approximation for
  inhomogeneous systems},\ }\href {https://doi.org/10.1103/PhysRevB.76.245113}
  {\bibfield  {journal} {\bibinfo  {journal} {Phys. Rev. B}\ }\textbf {\bibinfo
  {volume} {76}},\ \bibinfo {pages} {245113} (\bibinfo {year}
  {2007})}\BibitemShut {NoStop}%
\bibitem [{\citenamefont {Zhang}\ \emph {et~al.}(1988)\citenamefont {Zhang},
  \citenamefont {Gros}, \citenamefont {Rice},\ and\ \citenamefont
  {Shiba}}]{FCZhang_1988}%
  \BibitemOpen
  \bibfield  {author} {\bibinfo {author} {\bibfnamefont {F.~C.}\ \bibnamefont
  {Zhang}}, \bibinfo {author} {\bibfnamefont {C.}~\bibnamefont {Gros}},
  \bibinfo {author} {\bibfnamefont {T.~M.}\ \bibnamefont {Rice}},\ and\
  \bibinfo {author} {\bibfnamefont {H.}~\bibnamefont {Shiba}},\ }\bibfield
  {title} {\bibinfo {title} {A renormalised hamiltonian approach to a resonant
  valence bond wavefunction},\ }\href
  {https://doi.org/10.1088/0953-2048/1/1/009} {\bibfield  {journal} {\bibinfo
  {journal} {Superconductor Science and Technology}\ }\textbf {\bibinfo
  {volume} {1}},\ \bibinfo {pages} {36} (\bibinfo {year} {1988})}\BibitemShut
  {NoStop}%
\bibitem [{\citenamefont {Ghosal}\ \emph {et~al.}(2000)\citenamefont {Ghosal},
  \citenamefont {Randeria},\ and\ \citenamefont
  {Trivedi}}]{PhysRevB.63.020505}%
  \BibitemOpen
  \bibfield  {author} {\bibinfo {author} {\bibfnamefont {A.}~\bibnamefont
  {Ghosal}}, \bibinfo {author} {\bibfnamefont {M.}~\bibnamefont {Randeria}},\
  and\ \bibinfo {author} {\bibfnamefont {N.}~\bibnamefont {Trivedi}},\
  }\bibfield  {title} {\bibinfo {title} {Spatial inhomogeneities in disordered
  d-wave superconductors},\ }\href {https://doi.org/10.1103/PhysRevB.63.020505}
  {\bibfield  {journal} {\bibinfo  {journal} {Phys. Rev. B}\ }\textbf {\bibinfo
  {volume} {63}},\ \bibinfo {pages} {020505} (\bibinfo {year}
  {2000})}\BibitemShut {NoStop}%
\bibitem [{\citenamefont {Chakraborty}\ and\ \citenamefont
  {Ghosal}(2014)}]{Chakraborty_2014}%
  \BibitemOpen
  \bibfield  {author} {\bibinfo {author} {\bibfnamefont {D.}~\bibnamefont
  {Chakraborty}}\ and\ \bibinfo {author} {\bibfnamefont {A.}~\bibnamefont
  {Ghosal}},\ }\bibfield  {title} {\bibinfo {title} {Fate of disorder-induced
  inhomogeneities in strongly correlated d-wave superconductors},\ }\href
  {https://doi.org/10.1088/1367-2630/16/10/103018} {\bibfield  {journal}
  {\bibinfo  {journal} {New Journal of Physics}\ }\textbf {\bibinfo {volume}
  {16}},\ \bibinfo {pages} {103018} (\bibinfo {year} {2014})}\BibitemShut
  {NoStop}%
\bibitem [{\citenamefont {Kang}\ \emph {et~al.}(2023)\citenamefont {Kang},
  \citenamefont {Zhang}, \citenamefont {Schierle}, \citenamefont {McCoy},
  \citenamefont {Li}, \citenamefont {Sutarto}, \citenamefont {Suter},
  \citenamefont {Prokscha}, \citenamefont {Salman}, \citenamefont {Weschke},
  \citenamefont {Cybart}, \citenamefont {Wei},\ and\ \citenamefont
  {Comin}}]{doi:10.1073/pnas.2302099120}%
  \BibitemOpen
  \bibfield  {author} {\bibinfo {author} {\bibfnamefont {M.}~\bibnamefont
  {Kang}}, \bibinfo {author} {\bibfnamefont {C.~C.}\ \bibnamefont {Zhang}},
  \bibinfo {author} {\bibfnamefont {E.}~\bibnamefont {Schierle}}, \bibinfo
  {author} {\bibfnamefont {S.}~\bibnamefont {McCoy}}, \bibinfo {author}
  {\bibfnamefont {J.}~\bibnamefont {Li}}, \bibinfo {author} {\bibfnamefont
  {R.}~\bibnamefont {Sutarto}}, \bibinfo {author} {\bibfnamefont
  {A.}~\bibnamefont {Suter}}, \bibinfo {author} {\bibfnamefont
  {T.}~\bibnamefont {Prokscha}}, \bibinfo {author} {\bibfnamefont
  {Z.}~\bibnamefont {Salman}}, \bibinfo {author} {\bibfnamefont
  {E.}~\bibnamefont {Weschke}}, \bibinfo {author} {\bibfnamefont
  {S.}~\bibnamefont {Cybart}}, \bibinfo {author} {\bibfnamefont {J.~Y.~T.}\
  \bibnamefont {Wei}},\ and\ \bibinfo {author} {\bibfnamefont {R.}~\bibnamefont
  {Comin}},\ }\bibfield  {title} {\bibinfo {title} {Discovery of charge order
  in a cuprate mott insulator},\ }\href
  {https://doi.org/10.1073/pnas.2302099120} {\bibfield  {journal} {\bibinfo
  {journal} {Proceedings of the National Academy of Sciences}\ }\textbf
  {\bibinfo {volume} {120}},\ \bibinfo {pages} {e2302099120} (\bibinfo {year}
  {2023})}\BibitemShut {NoStop}%
\bibitem [{\citenamefont {da~Silva~Neto}\ \emph {et~al.}(2014)\citenamefont
  {da~Silva~Neto}, \citenamefont {Aynajian}, \citenamefont {Frano},
  \citenamefont {Comin}, \citenamefont {Schierle}, \citenamefont {Weschke},
  \citenamefont {Gyenis}, \citenamefont {Wen}, \citenamefont {Schneeloch},
  \citenamefont {Xu}, \citenamefont {Ono}, \citenamefont {Gu}, \citenamefont
  {Tacon},\ and\ \citenamefont {Yazdani}}]{doi:10.1126/science.1243479}%
  \BibitemOpen
  \bibfield  {author} {\bibinfo {author} {\bibfnamefont {E.~H.}\ \bibnamefont
  {da~Silva~Neto}}, \bibinfo {author} {\bibfnamefont {P.}~\bibnamefont
  {Aynajian}}, \bibinfo {author} {\bibfnamefont {A.}~\bibnamefont {Frano}},
  \bibinfo {author} {\bibfnamefont {R.}~\bibnamefont {Comin}}, \bibinfo
  {author} {\bibfnamefont {E.}~\bibnamefont {Schierle}}, \bibinfo {author}
  {\bibfnamefont {E.}~\bibnamefont {Weschke}}, \bibinfo {author} {\bibfnamefont
  {A.}~\bibnamefont {Gyenis}}, \bibinfo {author} {\bibfnamefont
  {J.}~\bibnamefont {Wen}}, \bibinfo {author} {\bibfnamefont {J.}~\bibnamefont
  {Schneeloch}}, \bibinfo {author} {\bibfnamefont {Z.}~\bibnamefont {Xu}},
  \bibinfo {author} {\bibfnamefont {S.}~\bibnamefont {Ono}}, \bibinfo {author}
  {\bibfnamefont {G.}~\bibnamefont {Gu}}, \bibinfo {author} {\bibfnamefont
  {M.~L.}\ \bibnamefont {Tacon}},\ and\ \bibinfo {author} {\bibfnamefont
  {A.}~\bibnamefont {Yazdani}},\ }\bibfield  {title} {\bibinfo {title}
  {Ubiquitous interplay between charge ordering and high-temperature
  superconductivity in cuprates},\ }\href
  {https://doi.org/10.1126/science.1243479} {\bibfield  {journal} {\bibinfo
  {journal} {Science}\ }\textbf {\bibinfo {volume} {343}},\ \bibinfo {pages}
  {393} (\bibinfo {year} {2014})}\BibitemShut {NoStop}%
\bibitem [{\citenamefont {Cai}\ \emph {et~al.}(2016)\citenamefont {Cai},
  \citenamefont {Ruan}, \citenamefont {Peng}, \citenamefont {Ye}, \citenamefont
  {Li}, \citenamefont {Hao}, \citenamefont {Zhou}, \citenamefont {Lee},\ and\
  \citenamefont {Wang}}]{Cai2016}%
  \BibitemOpen
  \bibfield  {author} {\bibinfo {author} {\bibfnamefont {P.}~\bibnamefont
  {Cai}}, \bibinfo {author} {\bibfnamefont {W.}~\bibnamefont {Ruan}}, \bibinfo
  {author} {\bibfnamefont {Y.}~\bibnamefont {Peng}}, \bibinfo {author}
  {\bibfnamefont {C.}~\bibnamefont {Ye}}, \bibinfo {author} {\bibfnamefont
  {X.}~\bibnamefont {Li}}, \bibinfo {author} {\bibfnamefont {Z.}~\bibnamefont
  {Hao}}, \bibinfo {author} {\bibfnamefont {X.}~\bibnamefont {Zhou}}, \bibinfo
  {author} {\bibfnamefont {D.-H.}\ \bibnamefont {Lee}},\ and\ \bibinfo {author}
  {\bibfnamefont {Y.}~\bibnamefont {Wang}},\ }\bibfield  {title} {\bibinfo
  {title} {Visualizing the evolution from the mott insulator to a
  charge-ordered insulator in lightly doped cuprates},\ }\href
  {https://doi.org/10.1038/nphys3840} {\bibfield  {journal} {\bibinfo
  {journal} {Nature Physics}\ }\textbf {\bibinfo {volume} {12}},\ \bibinfo
  {pages} {1047} (\bibinfo {year} {2016})}\BibitemShut {NoStop}%
\bibitem [{\citenamefont {Choubey}\ \emph {et~al.}(2017)\citenamefont
  {Choubey}, \citenamefont {Tu}, \citenamefont {Lee},\ and\ \citenamefont
  {Hirschfeld}}]{choubey2017incommensurate}%
  \BibitemOpen
  \bibfield  {author} {\bibinfo {author} {\bibfnamefont {P.}~\bibnamefont
  {Choubey}}, \bibinfo {author} {\bibfnamefont {W.-L.}\ \bibnamefont {Tu}},
  \bibinfo {author} {\bibfnamefont {T.-K.}\ \bibnamefont {Lee}},\ and\ \bibinfo
  {author} {\bibfnamefont {P.~J.}\ \bibnamefont {Hirschfeld}},\ }\bibfield
  {title} {\bibinfo {title} {Incommensurate charge ordered states in the
  $t\text{--}t^{\prime}\text{--}j$ model},\ }\href
  {https://doi.org/10.1088/1367-2630/19/1/013028} {\bibfield  {journal}
  {\bibinfo  {journal} {New Journal of Physics}\ }\textbf {\bibinfo {volume}
  {19}},\ \bibinfo {pages} {013028} (\bibinfo {year} {2017})}\BibitemShut
  {NoStop}%
\bibitem [{\citenamefont {Fukushima}(2008)}]{PhysRevB.78.115105}%
  \BibitemOpen
  \bibfield  {author} {\bibinfo {author} {\bibfnamefont {N.}~\bibnamefont
  {Fukushima}},\ }\bibfield  {title} {\bibinfo {title} {Grand canonical
  gutzwiller approximation for magnetic inhomogeneous systems},\ }\href
  {https://doi.org/10.1103/PhysRevB.78.115105} {\bibfield  {journal} {\bibinfo
  {journal} {Phys. Rev. B}\ }\textbf {\bibinfo {volume} {78}},\ \bibinfo
  {pages} {115105} (\bibinfo {year} {2008})}\BibitemShut {NoStop}%
\bibitem [{\citenamefont {Chakraborty}\ \emph {et~al.}(2017)\citenamefont
  {Chakraborty}, \citenamefont {Kaushal},\ and\ \citenamefont
  {Ghosal}}]{PhysRevB.96.134518}%
  \BibitemOpen
  \bibfield  {author} {\bibinfo {author} {\bibfnamefont {D.}~\bibnamefont
  {Chakraborty}}, \bibinfo {author} {\bibfnamefont {N.}~\bibnamefont
  {Kaushal}},\ and\ \bibinfo {author} {\bibfnamefont {A.}~\bibnamefont
  {Ghosal}},\ }\bibfield  {title} {\bibinfo {title} {Pairing theory for
  strongly correlated $d$-wave superconductors},\ }\href
  {https://doi.org/10.1103/PhysRevB.96.134518} {\bibfield  {journal} {\bibinfo
  {journal} {Phys. Rev. B}\ }\textbf {\bibinfo {volume} {96}},\ \bibinfo
  {pages} {134518} (\bibinfo {year} {2017})}\BibitemShut {NoStop}%
\bibitem [{\citenamefont {Hamidian}\ \emph
  {et~al.}(2016{\natexlab{b}})\citenamefont {Hamidian}, \citenamefont {Edkins},
  \citenamefont {Joo}, \citenamefont {Kostin}, \citenamefont {Eisaki},
  \citenamefont {Uchida}, \citenamefont {Lawler}, \citenamefont {Kim},
  \citenamefont {Mackenzie}, \citenamefont {Fujita}, \citenamefont {Lee},\ and\
  \citenamefont {Davis}}]{Hamidian2016}%
  \BibitemOpen
  \bibfield  {author} {\bibinfo {author} {\bibfnamefont {M.~H.}\ \bibnamefont
  {Hamidian}}, \bibinfo {author} {\bibfnamefont {S.~D.}\ \bibnamefont
  {Edkins}}, \bibinfo {author} {\bibfnamefont {S.~H.}\ \bibnamefont {Joo}},
  \bibinfo {author} {\bibfnamefont {A.}~\bibnamefont {Kostin}}, \bibinfo
  {author} {\bibfnamefont {H.}~\bibnamefont {Eisaki}}, \bibinfo {author}
  {\bibfnamefont {S.}~\bibnamefont {Uchida}}, \bibinfo {author} {\bibfnamefont
  {M.~J.}\ \bibnamefont {Lawler}}, \bibinfo {author} {\bibfnamefont {E.-A.}\
  \bibnamefont {Kim}}, \bibinfo {author} {\bibfnamefont {A.~P.}\ \bibnamefont
  {Mackenzie}}, \bibinfo {author} {\bibfnamefont {K.}~\bibnamefont {Fujita}},
  \bibinfo {author} {\bibfnamefont {J.}~\bibnamefont {Lee}},\ and\ \bibinfo
  {author} {\bibfnamefont {J.~C.~S.}\ \bibnamefont {Davis}},\ }\bibfield
  {title} {\bibinfo {title} {Detection of a cooper-pair density wave in
  $\text{Bi}_{2}\text{Sr}_{2}\text{Ca}\text{Cu}_{2}\text{O}_{8+x}$},\ }\href
  {https://doi.org/10.1038/nature17411} {\bibfield  {journal} {\bibinfo
  {journal} {Nature}\ }\textbf {\bibinfo {volume} {532}},\ \bibinfo {pages}
  {343} (\bibinfo {year} {2016}{\natexlab{b}})}\BibitemShut {NoStop}%
\bibitem [{\citenamefont {Achkar}\ \emph {et~al.}(2016)\citenamefont {Achkar},
  \citenamefont {He}, \citenamefont {Sutarto}, \citenamefont {McMahon},
  \citenamefont {Zwiebler}, \citenamefont {H{\"u}cker}, \citenamefont {Gu},
  \citenamefont {Liang}, \citenamefont {Bonn}, \citenamefont {Hardy},
  \citenamefont {Geck},\ and\ \citenamefont {Hawthorn}}]{Achkar2016}%
  \BibitemOpen
  \bibfield  {author} {\bibinfo {author} {\bibfnamefont {A.~J.}\ \bibnamefont
  {Achkar}}, \bibinfo {author} {\bibfnamefont {F.}~\bibnamefont {He}}, \bibinfo
  {author} {\bibfnamefont {R.}~\bibnamefont {Sutarto}}, \bibinfo {author}
  {\bibfnamefont {C.}~\bibnamefont {McMahon}}, \bibinfo {author} {\bibfnamefont
  {M.}~\bibnamefont {Zwiebler}}, \bibinfo {author} {\bibfnamefont
  {M.}~\bibnamefont {H{\"u}cker}}, \bibinfo {author} {\bibfnamefont {G.~D.}\
  \bibnamefont {Gu}}, \bibinfo {author} {\bibfnamefont {R.}~\bibnamefont
  {Liang}}, \bibinfo {author} {\bibfnamefont {D.~A.}\ \bibnamefont {Bonn}},
  \bibinfo {author} {\bibfnamefont {W.~N.}\ \bibnamefont {Hardy}}, \bibinfo
  {author} {\bibfnamefont {J.}~\bibnamefont {Geck}},\ and\ \bibinfo {author}
  {\bibfnamefont {D.~G.}\ \bibnamefont {Hawthorn}},\ }\bibfield  {title}
  {\bibinfo {title} {Orbital symmetry of charge-density-wave order in
  $\text{La}_{1.875}\text{Ba}_{0.125}\text{Cu}\text{O}_{4}$ and
  $\text{YBa}_{2}\text{Cu}_{3}\text{O}_{6.67}$},\ }\href
  {https://doi.org/10.1038/nmat4568} {\bibfield  {journal} {\bibinfo  {journal}
  {Nature Materials}\ }\textbf {\bibinfo {volume} {15}},\ \bibinfo {pages}
  {616} (\bibinfo {year} {2016})}\BibitemShut {NoStop}%
\bibitem [{\citenamefont {McMahon}\ \emph {et~al.}(2020)\citenamefont
  {McMahon}, \citenamefont {Achkar}, \citenamefont {da~Silva~Neto},
  \citenamefont {Djianto}, \citenamefont {Menard}, \citenamefont {He},
  \citenamefont {Sutarto}, \citenamefont {Comin}, \citenamefont {Liang},
  \citenamefont {Bonn}, \citenamefont {Hardy}, \citenamefont {Damascelli},\
  and\ \citenamefont {Hawthorn}}]{McMahon}%
  \BibitemOpen
  \bibfield  {author} {\bibinfo {author} {\bibfnamefont {C.}~\bibnamefont
  {McMahon}}, \bibinfo {author} {\bibfnamefont {A.~J.}\ \bibnamefont {Achkar}},
  \bibinfo {author} {\bibfnamefont {E.~H.}\ \bibnamefont {da~Silva~Neto}},
  \bibinfo {author} {\bibfnamefont {I.}~\bibnamefont {Djianto}}, \bibinfo
  {author} {\bibfnamefont {J.}~\bibnamefont {Menard}}, \bibinfo {author}
  {\bibfnamefont {F.}~\bibnamefont {He}}, \bibinfo {author} {\bibfnamefont
  {R.}~\bibnamefont {Sutarto}}, \bibinfo {author} {\bibfnamefont
  {R.}~\bibnamefont {Comin}}, \bibinfo {author} {\bibfnamefont
  {R.}~\bibnamefont {Liang}}, \bibinfo {author} {\bibfnamefont {D.~A.}\
  \bibnamefont {Bonn}}, \bibinfo {author} {\bibfnamefont {W.~N.}\ \bibnamefont
  {Hardy}}, \bibinfo {author} {\bibfnamefont {A.}~\bibnamefont {Damascelli}},\
  and\ \bibinfo {author} {\bibfnamefont {D.~G.}\ \bibnamefont {Hawthorn}},\
  }\bibfield  {title} {\bibinfo {title} {Orbital symmetries of charge density
  wave order in $\text{YBa}_{2}\text{Cu}_{3}\text{O}_{6+x}$},\ }\href
  {https://doi.org/10.1126/sciadv.aay0345} {\bibfield  {journal} {\bibinfo
  {journal} {Science Advances}\ }\textbf {\bibinfo {volume} {6}},\ \bibinfo
  {pages} {eaay0345} (\bibinfo {year} {2020})}\BibitemShut {NoStop}%
\bibitem [{\citenamefont {Gongora}\ \emph {et~al.}(2020)\citenamefont
  {Gongora}, \citenamefont {Xu}, \citenamefont {Perry}, \citenamefont {Okoye},
  \citenamefont {Riley}, \citenamefont {Reyes}, \citenamefont {Morgan},\ and\
  \citenamefont {Brown}}]{doi:10.1126/sciadv.aaz1708}%
  \BibitemOpen
  \bibfield  {author} {\bibinfo {author} {\bibfnamefont {A.~E.}\ \bibnamefont
  {Gongora}}, \bibinfo {author} {\bibfnamefont {B.}~\bibnamefont {Xu}},
  \bibinfo {author} {\bibfnamefont {W.}~\bibnamefont {Perry}}, \bibinfo
  {author} {\bibfnamefont {C.}~\bibnamefont {Okoye}}, \bibinfo {author}
  {\bibfnamefont {P.}~\bibnamefont {Riley}}, \bibinfo {author} {\bibfnamefont
  {K.~G.}\ \bibnamefont {Reyes}}, \bibinfo {author} {\bibfnamefont {E.~F.}\
  \bibnamefont {Morgan}},\ and\ \bibinfo {author} {\bibfnamefont {K.~A.}\
  \bibnamefont {Brown}},\ }\bibfield  {title} {\bibinfo {title} {A bayesian
  experimental autonomous researcher for mechanical design},\ }\href
  {https://doi.org/10.1126/sciadv.aaz1708} {\bibfield  {journal} {\bibinfo
  {journal} {Science Advances}\ }\textbf {\bibinfo {volume} {6}},\ \bibinfo
  {pages} {eaaz1708} (\bibinfo {year} {2020})}\BibitemShut {NoStop}%
\bibitem [{\citenamefont {Hess}\ \emph {et~al.}(1991)\citenamefont {Hess},
  \citenamefont {Robinson},\ and\ \citenamefont {Waszczak}}]{HESS1991422}%
  \BibitemOpen
  \bibfield  {author} {\bibinfo {author} {\bibfnamefont {H.}~\bibnamefont
  {Hess}}, \bibinfo {author} {\bibfnamefont {R.}~\bibnamefont {Robinson}},\
  and\ \bibinfo {author} {\bibfnamefont {J.}~\bibnamefont {Waszczak}},\
  }\bibfield  {title} {\bibinfo {title} {\text{STM} spectroscopy of vortex
  cores and the flux lattice},\ }\href
  {https://doi.org/https://doi.org/10.1016/0921-4526(91)90262-D} {\bibfield
  {journal} {\bibinfo  {journal} {Physica B: Condensed Matter}\ }\textbf
  {\bibinfo {volume} {169}},\ \bibinfo {pages} {422} (\bibinfo {year}
  {1991})}\BibitemShut {NoStop}%
\bibitem [{\citenamefont {Caroli}\ \emph {et~al.}(1964)\citenamefont {Caroli},
  \citenamefont {{De Gennes}},\ and\ \citenamefont {Matricon}}]{CAROLI1964307}%
  \BibitemOpen
  \bibfield  {author} {\bibinfo {author} {\bibfnamefont {C.}~\bibnamefont
  {Caroli}}, \bibinfo {author} {\bibfnamefont {P.}~\bibnamefont {{De
  Gennes}}},\ and\ \bibinfo {author} {\bibfnamefont {J.}~\bibnamefont
  {Matricon}},\ }\bibfield  {title} {\bibinfo {title} {Bound fermion states on
  a vortex line in a type ii superconductor},\ }\href
  {https://doi.org/https://doi.org/10.1016/0031-9163(64)90375-0} {\bibfield
  {journal} {\bibinfo  {journal} {Physics Letters}\ }\textbf {\bibinfo {volume}
  {9}},\ \bibinfo {pages} {307} (\bibinfo {year} {1964})}\BibitemShut {NoStop}%
\bibitem [{\citenamefont {Maggio-Aprile}\ \emph {et~al.}(1995)\citenamefont
  {Maggio-Aprile}, \citenamefont {Renner}, \citenamefont {Erb}, \citenamefont
  {Walker},\ and\ \citenamefont {Fischer}}]{PhysRevLett.75.2754}%
  \BibitemOpen
  \bibfield  {author} {\bibinfo {author} {\bibfnamefont {I.}~\bibnamefont
  {Maggio-Aprile}}, \bibinfo {author} {\bibfnamefont {C.}~\bibnamefont
  {Renner}}, \bibinfo {author} {\bibfnamefont {A.}~\bibnamefont {Erb}},
  \bibinfo {author} {\bibfnamefont {E.}~\bibnamefont {Walker}},\ and\ \bibinfo
  {author} {\bibfnamefont {O.}~\bibnamefont {Fischer}},\ }\bibfield  {title}
  {\bibinfo {title} {Direct vortex lattice imaging and tunneling spectroscopy
  of flux lines on $\text{YBa}_{2}\text{Cu}_{3}\text{O}_{7-\delta}$},\ }\href
  {https://doi.org/10.1103/PhysRevLett.75.2754} {\bibfield  {journal} {\bibinfo
   {journal} {Phys. Rev. Lett.}\ }\textbf {\bibinfo {volume} {75}},\ \bibinfo
  {pages} {2754} (\bibinfo {year} {1995})}\BibitemShut {NoStop}%
\bibitem [{\citenamefont {Hoogenboom}\ \emph {et~al.}(2000)\citenamefont
  {Hoogenboom}, \citenamefont {Renner}, \citenamefont {Revaz}, \citenamefont
  {Maggio-Aprile},\ and\ \citenamefont {Fischer}}]{HOOGENBOOM2000440}%
  \BibitemOpen
  \bibfield  {author} {\bibinfo {author} {\bibfnamefont {B.}~\bibnamefont
  {Hoogenboom}}, \bibinfo {author} {\bibfnamefont {C.}~\bibnamefont {Renner}},
  \bibinfo {author} {\bibfnamefont {B.}~\bibnamefont {Revaz}}, \bibinfo
  {author} {\bibfnamefont {I.}~\bibnamefont {Maggio-Aprile}},\ and\ \bibinfo
  {author} {\bibfnamefont {O.}~\bibnamefont {Fischer}},\ }\bibfield  {title}
  {\bibinfo {title} {Low-energy structures in vortex core tunneling spectra in
  $\text{Bi}_{2}\text{Sr}_{2}\text{Ca}\text{Cu}_{2}\text{O}_{8+\delta}$},\
  }\href {https://doi.org/https://doi.org/10.1016/S0921-4534(99)00720-0}
  {\bibfield  {journal} {\bibinfo  {journal} {Physica C: Superconductivity}\
  }\textbf {\bibinfo {volume} {332}},\ \bibinfo {pages} {440} (\bibinfo {year}
  {2000})}\BibitemShut {NoStop}%
\bibitem [{\citenamefont {Pan}\ \emph {et~al.}(2000)\citenamefont {Pan},
  \citenamefont {Hudson}, \citenamefont {Gupta}, \citenamefont {Ng},
  \citenamefont {Eisaki}, \citenamefont {Uchida},\ and\ \citenamefont
  {Davis}}]{PhysRevLett.85.1536}%
  \BibitemOpen
  \bibfield  {author} {\bibinfo {author} {\bibfnamefont {S.~H.}\ \bibnamefont
  {Pan}}, \bibinfo {author} {\bibfnamefont {E.~W.}\ \bibnamefont {Hudson}},
  \bibinfo {author} {\bibfnamefont {A.~K.}\ \bibnamefont {Gupta}}, \bibinfo
  {author} {\bibfnamefont {K.-W.}\ \bibnamefont {Ng}}, \bibinfo {author}
  {\bibfnamefont {H.}~\bibnamefont {Eisaki}}, \bibinfo {author} {\bibfnamefont
  {S.}~\bibnamefont {Uchida}},\ and\ \bibinfo {author} {\bibfnamefont {J.~C.}\
  \bibnamefont {Davis}},\ }\bibfield  {title} {\bibinfo {title} {\text{STM}
  studies of the electronic structure of vortex cores in
  $\text{Bi}_{2}\text{Sr}_{2}\text{Ca}\text{Cu}_{2}\text{O}_{8+\delta}$},\
  }\href {https://doi.org/10.1103/PhysRevLett.85.1536} {\bibfield  {journal}
  {\bibinfo  {journal} {Phys. Rev. Lett.}\ }\textbf {\bibinfo {volume} {85}},\
  \bibinfo {pages} {1536} (\bibinfo {year} {2000})}\BibitemShut {NoStop}%
\bibitem [{\citenamefont {Gazdi\ifmmode~\acute{c}\else \'{c}\fi{}}\ \emph
  {et~al.}(2021)\citenamefont {Gazdi\ifmmode~\acute{c}\else \'{c}\fi{}},
  \citenamefont {Maggio-Aprile}, \citenamefont {Gu},\ and\ \citenamefont
  {Renner}}]{PhysRevX.11.031040}%
  \BibitemOpen
  \bibfield  {author} {\bibinfo {author} {\bibfnamefont {T.}~\bibnamefont
  {Gazdi\ifmmode~\acute{c}\else \'{c}\fi{}}}, \bibinfo {author} {\bibfnamefont
  {I.}~\bibnamefont {Maggio-Aprile}}, \bibinfo {author} {\bibfnamefont
  {G.}~\bibnamefont {Gu}},\ and\ \bibinfo {author} {\bibfnamefont
  {C.}~\bibnamefont {Renner}},\ }\bibfield  {title} {\bibinfo {title}
  {Wang-macdonald $d$-wave vortex cores observed in heavily overdoped
  $\text{Bi}_{2}\text{Sr}_{2}\text{Ca}\text{Cu}_{2}\text{O}_{8+\delta}$},\
  }\href {https://doi.org/10.1103/PhysRevX.11.031040} {\bibfield  {journal}
  {\bibinfo  {journal} {Phys. Rev. X}\ }\textbf {\bibinfo {volume} {11}},\
  \bibinfo {pages} {031040} (\bibinfo {year} {2021})}\BibitemShut {NoStop}%
\bibitem [{\citenamefont {Chang}\ \emph {et~al.}(2012)\citenamefont {Chang},
  \citenamefont {Blackburn}, \citenamefont {Holmes}, \citenamefont
  {Christensen}, \citenamefont {Larsen}, \citenamefont {Mesot}, \citenamefont
  {Liang}, \citenamefont {Bonn}, \citenamefont {Hardy}, \citenamefont
  {Watenphul}, \citenamefont {Zimmermann}, \citenamefont {Forgan},\ and\
  \citenamefont {Hayden}}]{Chang2012}%
  \BibitemOpen
  \bibfield  {author} {\bibinfo {author} {\bibfnamefont {J.}~\bibnamefont
  {Chang}}, \bibinfo {author} {\bibfnamefont {E.}~\bibnamefont {Blackburn}},
  \bibinfo {author} {\bibfnamefont {A.~T.}\ \bibnamefont {Holmes}}, \bibinfo
  {author} {\bibfnamefont {N.~B.}\ \bibnamefont {Christensen}}, \bibinfo
  {author} {\bibfnamefont {J.}~\bibnamefont {Larsen}}, \bibinfo {author}
  {\bibfnamefont {J.}~\bibnamefont {Mesot}}, \bibinfo {author} {\bibfnamefont
  {R.}~\bibnamefont {Liang}}, \bibinfo {author} {\bibfnamefont {D.~A.}\
  \bibnamefont {Bonn}}, \bibinfo {author} {\bibfnamefont {W.~N.}\ \bibnamefont
  {Hardy}}, \bibinfo {author} {\bibfnamefont {A.}~\bibnamefont {Watenphul}},
  \bibinfo {author} {\bibfnamefont {M.~v.}\ \bibnamefont {Zimmermann}},
  \bibinfo {author} {\bibfnamefont {E.~M.}\ \bibnamefont {Forgan}},\ and\
  \bibinfo {author} {\bibfnamefont {S.~M.}\ \bibnamefont {Hayden}},\ }\bibfield
   {title} {\bibinfo {title} {Direct observation of competition between
  superconductivity and charge density wave order in
  $\text{YBa}_{2}\text{Cu}_{3}\text{O}_{6.67}$},\ }\href
  {https://doi.org/10.1038/nphys2456} {\bibfield  {journal} {\bibinfo
  {journal} {Nature Physics}\ }\textbf {\bibinfo {volume} {8}},\ \bibinfo
  {pages} {871} (\bibinfo {year} {2012})}\BibitemShut {NoStop}%
\bibitem [{\citenamefont {Wise}\ \emph {et~al.}(2008)\citenamefont {Wise},
  \citenamefont {Boyer}, \citenamefont {Chatterjee}, \citenamefont {Kondo},
  \citenamefont {Takeuchi}, \citenamefont {Ikuta}, \citenamefont {Wang},\ and\
  \citenamefont {Hudson}}]{Wise2008}%
  \BibitemOpen
  \bibfield  {author} {\bibinfo {author} {\bibfnamefont {W.~D.}\ \bibnamefont
  {Wise}}, \bibinfo {author} {\bibfnamefont {M.~C.}\ \bibnamefont {Boyer}},
  \bibinfo {author} {\bibfnamefont {K.}~\bibnamefont {Chatterjee}}, \bibinfo
  {author} {\bibfnamefont {T.}~\bibnamefont {Kondo}}, \bibinfo {author}
  {\bibfnamefont {T.}~\bibnamefont {Takeuchi}}, \bibinfo {author}
  {\bibfnamefont {H.}~\bibnamefont {Ikuta}}, \bibinfo {author} {\bibfnamefont
  {Y.}~\bibnamefont {Wang}},\ and\ \bibinfo {author} {\bibfnamefont {E.~W.}\
  \bibnamefont {Hudson}},\ }\bibfield  {title} {\bibinfo {title}
  {Charge-density-wave origin of cuprate checkerboard visualized by scanning
  tunnelling microscopy},\ }\href {https://doi.org/10.1038/nphys1021}
  {\bibfield  {journal} {\bibinfo  {journal} {Nature Physics}\ }\textbf
  {\bibinfo {volume} {4}},\ \bibinfo {pages} {696} (\bibinfo {year}
  {2008})}\BibitemShut {NoStop}%
\bibitem [{\citenamefont {Frano}\ \emph {et~al.}(2020)\citenamefont {Frano},
  \citenamefont {Blanco-Canosa}, \citenamefont {Keimer},\ and\ \citenamefont
  {Birgeneau}}]{Frano_2020}%
  \BibitemOpen
  \bibfield  {author} {\bibinfo {author} {\bibfnamefont {A.}~\bibnamefont
  {Frano}}, \bibinfo {author} {\bibfnamefont {S.}~\bibnamefont
  {Blanco-Canosa}}, \bibinfo {author} {\bibfnamefont {B.}~\bibnamefont
  {Keimer}},\ and\ \bibinfo {author} {\bibfnamefont {R.~J.}\ \bibnamefont
  {Birgeneau}},\ }\bibfield  {title} {\bibinfo {title} {Charge ordering in
  superconducting copper oxides},\ }\href
  {https://doi.org/10.1088/1361-648X/ab6140} {\bibfield  {journal} {\bibinfo
  {journal} {Journal of Physics: Condensed Matter}\ }\textbf {\bibinfo {volume}
  {32}},\ \bibinfo {pages} {374005} (\bibinfo {year} {2020})}\BibitemShut
  {NoStop}%
\end{thebibliography}%

\end{document}